\documentclass[11pt]{article}

\usepackage{amsmath,amssymb}
\usepackage{graphicx}
\usepackage{psfrag}

\newcommand{\R}{\mathbb{R}}
\newcommand{\C}{\mathbb{C}}

\newcommand{\g}{\gamma}
\renewcommand{\a}{\alpha}
\newcommand{\A}{{\cal A}}

\newcommand{\Abar}{\overline{\cal A}}

\renewcommand{\S}{{\cal S}}
\newcommand{\Z}{\mathbb{Z}}
\newcommand{\N}{\mathbb{N}}
\newcommand{\h}{{\cal H}}
\newcommand{\scal}[2]{\langle #1| #2\rangle}

\newcommand{\cyl}{{\rm Cyl}}

\newcommand{\scripta}{\mathfrak{A}}

\newcommand{\ot}{\otimes}

\newcommand{\bG}{\boldsymbol{G}}

\newcommand{\bld}[1]{\boldsymbol{#1}}
\newcommand{\tl}[1]{\tilde{#1}}
\newcommand{\tll}{\tilde{l}}

\newcommand{\tame}{\tilde{\mathfrak{L}}}

\newcounter{mnotecount}[section]

\newtheorem{lm}{Lemma}
\newtheorem{df}{Definition}
\newtheorem{thr}[lm]{Theorem}

\numberwithin{equation}{section}
\numberwithin{lm}{section}
\numberwithin{chr}{section}
\numberwithin{df}{section}

\begin{document}

\title{Hilbert space built over connections with a non-compact structure group}
\author{ Andrzej Oko{\l}\'ow}
\date{June 8, 2004}

\maketitle
\begin{center}
{\it  Instytut  Fizyki Teoretycznej, Uniwersytet
Warszawski,\\ ul. Ho\.{z}a 69, 00-681 Warsaw, Poland;\\
oko@fuw.edu.pl}
\end{center}
\medskip

\begin{abstract}
Quantization of general relativity in terms of $SL(2,\C)$-connections (i.e. in terms of the complex Ashtekar variables) is technically difficult because of the non-compactness of $SL(2,\C)$. The difficulties concern the construction of a diffeomorphism invariant Hilbert space structure on the space of cylindrical functions of the connections. We present here a 'toy' model of such a Hilbert space built over connections whose structure group is the group of real numbers. We show that in the case of any Hilbert space built analogously over connections with any non-compact structure group (this includes some models presented in the literature) there exists an obstacle which does not allow to define a $*$-representation of cylindrical functions on the Hilbert space by the multiplication map which is the only known way to define a diffeomorphism invariant representation of the functions.

\end{abstract}


\section{Introduction}

Canonical quantization of diffeomorphism invariant theories of connections\footnote{A diffeomorphism invariant theory of connections means  a theory in a Hamiltonian form such that $(i)$ its configuration space is a space of connections on a principal bundle $P(\Sigma,G)$, where $\Sigma$ is a base manifold, and $G$ is a Lie group (i.e. the structure group of the bundle and the connections), $(ii)$ there exist Gauss and vector constraints imposed on the phase space which ensures, respectively, gauge and diffeomorphism invariance of the theory \cite{cq-diff}.} is nowadays a well-established procedure \cite{cq-diff}. However, currently it can be applied only to theories of connections with a {\em compact} structure group. This situation is not satisfactory, mostly because general relativity (GR) which was the main motivation for creating the procedure, can be naturally seen as a theory of connections with a {\em non-compact} structure group (in the Ashtekar formulation \cite{a-var} the group is $SL(2,\C)$, that is a double covering of the proper Lorentz group). Thus  we are currently unable to apply the procedure {\em directly} to GR and therefore we are forced to formulate and quantize GR as a theory of $SU(2)$-connections. Although that is performable we cannot neglect some worrisome features of the resulting quantum theory such $(i)$ the lack of Lorentz symmetry which is broken in a non-natural way, $(ii)$ some operators (e.g. area operators) possessing clear geometrical/physical interpretation depend on the so-called Immirzi parameter \cite{immirzi} which at the classical level seems to be meaningless since it labels some canonical transformations of the phase space of GR; one of the consequences of this fact is that the entropy of a black hole derived in the framework of the quantum theory also depends on the parameter \cite{bhe}; $(iii)$ the Hamiltonian constraint expressed in terms of $SU(2)$-connections is much more complicated than the same constraint expressed in terms of $SL(2,\C)$ ones (see e.g. \cite{rev}), hence to obtain the corresponding Hamiltonian operator one has to apply quite a sophisticated regularization procedure \cite{qsd}; one hopes that the application of $SL(2,\C)$-connections can simplify this procedure. On the other hand, at present there are no known 'no-go' theorems which would claim that the application of canonical quantization (in the form presented in \cite{cq-diff}) to a theory of connections with a non-compact structure group is impossible, and difficulties with the application  seem to be rather technical. 

Thus any attempt to extend the applicability of  the canonical quantization procedure to the non-compact case is well motivated and not worthless. 

The present paper is devoted to a description of such an attempt which consists in a slight modification of the framework known from  the compact case. We will show, however, that the modification is not sufficient to obtain a satisfactory result. Despite of this it seems to us that the attempt should be presented, because understanding 'negative' results can also be helpful while looking for a solution of a problem. Moreover, the attempt was independently presented in the literature \cite{freidel}, but a deficiency of the result was overlooked there. Another construction which suffers from the deficiency can be found in \cite{liv}.

 Let us give now a description of the difficulties which one encounters trying to extend the quantization procedure to the non-compact case. After this we will present an outline of the attempt mentioned above. 

\subsection{Why are we not able to apply the quantization procedure to the non-compact case?}

Canonical quantization \cite{cq-diff} requires that we choose a set of functions on the phase space of the theory under quantization (these functions are called {\em elementary variables}) and then find a representation of the functions on a Hilbert space. Moreover, diffeomorphism invariance of the theory imposes some invariance conditions on the set of elementary variables, its representations and the Hilbert space --- the natural action of diffeomorphisms of the 'spatial' manifold on the phase space of the theory can be lifted to an action on functions on the phase space; thus we require the set of elementary variables, the representation and the scalar product on the Hilbert space to be invariant with respect to the action of diffeomorphisms (see e.g. \cite{diff-inv}).

Natural candidates for elementary variables for a diffeomorphism invariant theory of connections are so called cylindrical and flux functions \cite{ai,area,acz} --- cylindrical functions depend on connections as configuration variable of the theory under quantization (each such a function depends on holonomies of the connections along a finite number of paths), while flux functions depend on the momentum variable conjugate to the connections (any flux function is defined as an integral of the momentum variable over a surface).  Thus the functions are built without using any additional structure like e.g. a background metric on the base manifold which would be an obstacle for diffeomorphism invariance of the set of the variables. 

The next step of the quantization procedure is an assignment of an operator $\hat{f}$ on a Hilbert space to the elementary variable $f$. We require this assignment to satisfy the following two conditions\footnote{For a comprehensive description of the assignment see \cite{cq-diff}.}:
\begin{equation}
\widehat{(\bar{f}\,)}=(\hat{f}\,)^* \ \ \ \text{and} \ \ \ [\hat{f},\hat{g}]=i\,\widehat{\{f,g\}},
\label{star-rep}
\end{equation}
where $\bar{f}$ is the complex conjugate of the function $f$, and $(\hat{f}\,)^*$ denotes the operator adjoint to $\hat{f}$. Thus the assignment $f\mapsto\hat{f}$ will be called a $*$-representation of elementary variables.
     
In practice, to construct a $*$-representation of cylindrical and flux functions  we repeat the standard procedure used in quantum mechanics --- we try to define the Hilbert space as a space of square-integrable functions with respect to a measure on the configuration space (that is on the space of connections). Then cylindrical functions are supposed to  be represented on the Hilbert space by multiplication of functions, and flux functions --- by some differential operators. We do not claim that it is necessary to proceed like that, but in practice this is the only known  way to obtain a diffeomorphism invariant $*$-representation of the elementary variables on a Hilbert space equipped with a diffeomorphism invariant scalar product. This strategy can be successfully performed in the compact case but it fails in the non-compact one. The reason is that we require $(i)$ the scalar product on the Hilbert space to be diffeomorphism invariant and $(ii)$ the scalar product to be given by a measure on the space of connections; consequently, we have to find a diffeomorphism invariant measure on the space of connections which is still unachievable in the non-compact case (we emphasize again that there are no known theorems which forbid the existence of such a measure, we just do not know how to find it). Let us show now why the construction of a diffeomorphism invariant measure used successfully in the compact case fails in the non-compact one. 

Consider a principal bundle $P(\Sigma,G)$, where $G$ is non-compact. 
Let $\{e_1,\ldots, e_N\}$ be paths (edges) in the base ('spatial') manifold $\Sigma$ which form a graph $\g$. Given a connection $A$, let $g_i$ ($i=1,2,\ldots N$) be an element of $G$ describing the holonomy of $A$ along the path $e_i$. Then, given a function $\psi:G^N\mapsto \C$, we define a cylindrical function {\em compatible} with the graph $\g$ as
\[
\Psi(A)=\psi(g_1,\ldots,g_N).
\]           
If $\psi$ is integrable with respect to a measure $d\mu$ on $G^N$ then we can assign:
\begin{equation}
\Psi \mapsto \omega(\Psi):=\int_{G^N}\, \psi \, d\mu \in \C. 
\label{int-fun}
\end{equation}
In the compact case, if e.g. $d\mu$ is the Haar measure on $G^N$ this assignment leads to a positive, diffeomorphism invariant  functional defined on the algebra of all cylindrical functions and by virtue of Riesz-Markov theorem defines the desired measure on the space of connections \cite{al-hoop, proj}. But in the non-compact case the resulting functional\footnote{Here we assume that the assignment does lead to a functional on the space of cylindrical functions which actually may not be true.} $\omega$ on the space of cylindrical functions  cannot simultaneously be {\em positive} and {\em diffeomorphism invariant}, as it is shown by the following Fleischhack-Lewandowski-Sahlmann argument \cite{priv,a-notes}. 

Consider a path $e$, its image $\phi(e)$ under a diffeomorphism $\phi$ of the base manifold and a cylindrical function
\[
\Psi(A)=\psi(g),
\]
where $g$ is a holonomy of $A$ along the path $e$, and $\psi$ is the characteristic function of a subset of $G$. Let $\Psi'$ be a function obtained from $\Psi$ by means of the diffeomorphism, i.e. 
\[
\Psi'(A)=\psi(g'),
\]
where $g'$ is a holonomy of $A$ along the path $\phi(e)$. Then $(\Psi-\Psi')^2\geq 0$ and 
\[
\omega((\Psi-\Psi')^2)=\omega(\Psi^2)+\omega(\Psi^{\prime 2})-2\omega(\Psi\Psi').
\]
Diffeomorphism invariance of $\omega$ means that $\omega(\Psi^2)=\omega(\Psi^{\prime 2})$, hence
\begin{multline*}
\omega((\Psi-\Psi')^2)=2\omega(\Psi^2)-2\omega(\Psi\Psi')=2\int_G \psi(g) \, d\mu(g)-\\ -2 \int_{G^2}\psi(g_1)\psi(g_2)\, d\mu'(g_1,g_2).
\end{multline*}
Now, if we assume that $(i)$ the measures $d\mu$ and $d\mu'$ are non-normalizable (which is natural in the non-compact case) and that, for example, $(ii)$ $d\mu'=d\mu\times d\mu$, then we can easily find a function $\psi$ corresponding to a set whose measure is large enough to ensure that $\omega((\Psi-\Psi')^2)<0$. We emphasize that this result can be obtained under an assumption much weaker than $(ii)$ (as it was noted in \cite{a-notes}, instead of $(ii)$ one can use the assumption \ref{<} of Lemma \ref{undef}); on the other hand it is difficult to avoid the assumption $(i)$. Thus we have to admit that we do not know any reasonable\footnote{Such a measure should be also gauge invariant; see the discussion below.}  measure on $G^N$ which would give us the desired functional $\omega$ on the space of connections and that, consequently, we cannot perform the second step of the canonical quantization procedure in the non-compact case.

We emphasize that we failed in our attempt to define the desired Hilbert space as {\em a space of square-integrable functions} with respect to a measure on the space of connections. But the Hilbert space that we are looking for does not have to be obtained in that way and we can suppose that a Hilbert space constructed in another way can be good enough for our purposes. Let us then describe a strategy\footnote{The strategy presented here is a modification of one considered by Abhay Ashtekar and  Jerzy Lewandowski \cite{priv}.} which will lead to a {\em diffeomorphism invariant} Hilbert space structure defined on some functions of connections with non-compact structure group.     

\subsection{Outline of a construction of a diffeomorphism invariant Hilbert space \label{outline}}

Following the compact case we are going to define a Hilbert space structure on a space of some cylindrical functions. Notice that in the compact case cylindrical functions play double role: they are $(i)$ elementary variables and $(ii)$ vectors of the Hilbert space on which a $*$-representation of the elementary variables is defined. However, we cannot assume that it will remain true in the non-compact case --- e.g. in quantum mechanics whose configuration space is non-compact  elementary variables (i.e. polynomials of the Cartesian coordinates on $\R^3$) do not belong to the Hilbert space $L^2(\R^3,dx)$. Thus since now until the discussion in Section \ref{disc} we will consider some  cylindrical functions merely as vectors in a Hilbert space (more precisely, as ingredients from which a Hilbert space will be built). Only in Section \ref{disc} we will try to answer the question what kind of cylindrical functions can be applied as elementary variables. After this remark let us give the outline introduced in the title of the present subsection.

Consider again a principal bundle $P(\Sigma,G)$ over a base manifold $\Sigma$ with a non-compact structure group. Given a graph $\g\subset\Sigma$ of $N$ edges, consider a set of {\em gauge invariant} cylindrical functions compatible with it (notice that the class of theories under consideration contains only gauge invariant ones). Every cylindrical function compatible with the graph $\g$   can be regarded as a function on $G^N$, consequently every gauge invariant cylindrical function compatible with $\g$  can be regarded as a function on some quotient space $G^N/\sim$, where the relation $\sim$ on $G^N$ is induced by the gauge transformations \cite{freidel}. Assume now that we know how to define a measure on the space  $G^N/\sim$ corresponding to any graph and  consider a linear space of all gauge invariant cylindrical functions, {\em square-integrable} with respect to appropriate measures. Then we can define a diffeomorphism invariant, positive definite scalar product on the space as follows:  
\begin{enumerate}
\item given two gauge invariant functions compatible with the {\em same} graph,  we define the scalar product between them by means of the integral with respect to the measure on the corresponding space $G^N/\sim$;
\item any two gauge invariant functions compatible with {\em distinct} graphs are mutually orthogonal.
\end{enumerate}

In the general case, i.e. when the group $G$ is non-commutative, the construction of measures on the spaces $G^N/\sim$ is a non-trivial task (however, such measures were found in \cite{freidel}). Therefore here we will restrict ourselves only to the commutative group $\R$ of real numbers (which is the simplest non-compact Lie group) and applying the strategy just described we will obtain as a result a  'toy' model of a Hilbert space over connections with a non-compact structure group. Because of commutativity of $\R$ every cylindrical function compatible with loops is gauge invariant\footnote{This is because in the case of connections with a commutative structure group the holonomy along any loop is gauge invariant.} and can be viewed as a function on $\R^n$ for some $n$. Thus $(i)$ instead of graphs embedded in $\Sigma$ we will use sets of loops and $(ii)$  the scalar product between cylindrical functions compatible with the same set of loops  will be defined by the Lebesgue measure on $\R^n$. 

Finally, proceeding as described above we will obtain a {\em diffeomorphism invariant positive definite} scalar product on the space of some gauge invariant functions of connections with the structure group $\R$. A completion of the space with respect to the norm provided by the scalar product will give us the desired Hilbert space. 

It turns out, however, that the Hilbert space does not solve our main problem because it is not clear how to define  any non-trivial $*$-rep\-re\-sen\-ta\-tion of cylindrical functions on the space\footnote{Here we mean some cylindrical functions treated as elementary variables.} --- the structure of the Hilbert space does not admit a $*$-representation of the functions defined by the multiplication map which is a standard way of proceeding (strictly speaking, the multiplication map can give us at most a representation of the functions, and  not a $*$-representation).  We once again emphasize that we do not have to use the multiplication map to get a $*$-representation of the cylindrical functions but,  in fact, $(i)$ the 'no-go' result concerning the multiplication map can be extended\footnote{This extension will be described in Subsection \ref{ext}.} to a wider class of representations of cylindrical functions on the Hilbert space \cite{a-notes} and $(ii)$ we do not know any other way to define a (diffeomorphism invariant) representation of cylindrical functions on the Hilbert space. Moreover, we will show that a wide class of modifications of the Hilbert space also does not admit a $*$-representation given by multiplication. Let us finally notice that the 'negative' results are not limited only to the 'toy' model --- they are also valid in the cases of $(i)$  a Hilbert space, constructed by Freidel and Livine \cite{freidel}, over connections  with a non-compact semi-simple structure group and $(ii)$ a Hilbert space of so called projected cylindrical functions due to Livine \cite{liv}. 

The plan of the paper is the following: Section 2 contains some basic facts concerning 'ingredients' which will be used in the construction of the Hilbert space; we will consider in turn: a group of holonomically equivalent loops (i.e. a hoop group), a space of generalized connections on a trivial bundle $\Sigma\times\R$, Schwarz functions on $\R^n$ and cylindrical functions defined by them. Section 3 is devoted to the construction of the 'toy' Hilbert space, and in Section 4 we will present and discuss the 'negative' results concerning $*$-representations of cylindrical functions on the Hilbert space.


\section{Preliminaries}

Let $\A$ be a space of smooth connections on a trivial bundle $P=\Sigma\times \R$, where $\Sigma$ is a real analytic, $d$-dimensional manifold ($d\geq 2$), and $\R$ is a group of real numbers with the group action provided by addition of the numbers. $\A$  will be called a space of $\R$-connections.

\subsection{Hoop groups}

As we noticed earlier commutativity of the group $\R$ implies that every (cylindrical) function of $\R$-connections which depend on holonomies along (a finite number of) loops is gauge invariant. Therefore  to the construction of the Hilbert space it is more convenient to  apply loops than graphs. 

Let $e$ be an oriented  piecewise analytic path embedded in $\Sigma$. We will denote by $e^{-1}$ a path obtained from $e$ by the change of the orientation. Given two oriented piecewise analytic paths $e_1,e_2$ such that $e_1$ ends at this point at which $e_2$ originates, we will denote by $e_2\circ e_1$ an oriented path obtained as a composition of the paths $e_1,e_2$.        

Following \cite{al-hoop} we introduce the notion of a hoop. Let $ \mathbb{L}_y$ be the set of all oriented piecewise analytic loops which originate and end at a point $y\in\Sigma$.  Two loops $l,l'\in \mathbb{L}_y$ will be said to be holonomically equivalent, $l\sim l'$, if and only if for every connection $A\in\A$:
\[
H(l,A)=H(l',A),
\]  
where $H(l,A)$ is a holonomy along the loop $l$ defined by the connection $A$ (for an explicit formula describing the holonomy see Appendix \ref{hol-app}). Denote by $\tll$ the equivalence class of loop $l$ and call it a {\em hoop}. The set of all hoops
\[
{\cal HG}:=\mathbb{L}_y/\sim
\] 
is an Abelian group called the {\em hoop group} with the group action given by
\[
\tll\circ\tll'=\widetilde{l\circ l'}.
\]

\subsubsection{Tame subgroups of $\cal HG$}

The definition of cylindrical functions on $\cal A$ will be based on some finitely generated subgroups of $\cal HG$ which we are going to introduce now. 

A subset ${\cal L}=\{l_1,\ldots,l_N\}$ of $\mathbb{L}_y$ is called a set of independent loops if and only if  each loop $l_I$ contains an open segment which is traversed only once and which is shared by any other loop at most at a finite number of points and $(ii)$ it does not contain any path of the form $e\circ e^{-1}$ where $e$ is a piecewise analytic path in $\Sigma$ \cite{proj}. 

Let ${\cal L}=\{ l_1,\ldots, l_n\}$ be a set of independent loops. A subgroup  $\tl{\cal L}$ of the hoop group $\cal HG$ generated by hoops $\{\tll_1,\ldots,\tll_n\}$ is said to be a {\em tame} subgroup of $\cal HG$ \cite{proj}. In the sequel we will also say that the group $\tl{\cal L}$ is generated by the loops ${\cal L}$  which is not too precise but convenient. Now we are going to establish some properties of tame subgroups of $\cal HG$ which will be useful while constructing the Hilbert space.  

We have an important lemma \cite{al-hoop}:
\begin{lm}
Given a set ${\cal L}=\{ l_1,\ldots, l_n\}$ of independent loops, for every $(x_1,\ldots,x_n)\in\R^n$ there exists a (smooth) connection $A\in\A$ such that:
\[
x_j=A(\tll_j).
\]   
\label{x-Al}
\end{lm}
Now it is easy to prove 
 \begin{lm}
Suppose that loops $l, {l'}$ are of the form 
\[
 l= k^{\eta_1}_{i_1}\circ\ldots\circ k^{\eta_N}_{i_N};\;\;\; {l'}= k^{\eta'_1}_{j_1}\circ\ldots\circ k^{\eta'_M}_{j_M},
\]
where $\{ k_1,\ldots, k_n\}$ is a set of independent loops,  $i_I,j_J\in\{1,\ldots,n\}$ and $\eta_I,\eta'_J\in\mathbb{Z}$. Then:
\[
\tll=\tilde{{l'}}\ \ \text{\rm if and only if} \ \  \sum_{\substack{K=1\\m=i_K}}^N\eta_K=\sum_{\substack{L=1\\m=j_L}}^M\eta'_L.
\] 
for all $m\in\{1,\ldots,n\}$. 
\label{power}
\end{lm}

Lemma \ref{x-Al} implies in particular that, given a set ${\cal L}=\{l_1,\ldots,l_n\}$ of independent loops, there exists a set $\{A_1,\ldots,A_n\}$ of $\R$-connections such that:
\[
A_i(\tll_j)=\delta_{ij}.
\] 
This allows us to construct a map $\rho_{\cal L}:\tl{\cal L}\rightarrow\R^n$ in the following way:
\begin{equation}
\rho_{\cal L}(\tilde{l}):=(A_1(\tilde{l}),\ldots,A_n(\tilde{l})).
\label{ln}
\end{equation}
In fact the map $\rho_{\cal L}$ is a group homomorphism --- for any two hoops $\tll_1,\tll_2\in{\cal L}$ we have  
\[
\rho_{\cal L}(\tilde{l}_1 \circ \tilde{l}_2)=\rho_{\cal L}(\tilde{l}_1)+\rho_{\cal L}(\tilde{l}_2).
\]
Moreover, for the generators $\{\tll_1,\ldots,\tll_n\}$ of $\tl{\cal L}$ we have 
\[
\rho_{\cal L}(\tll_i)=\varepsilon_i,
\]
where $(\varepsilon_1,\ldots,\varepsilon_n)$ is the canonical basis of $\R^n$. Using Lemma \ref{power} we easily get 
\begin{lm}
The map $\rho_{\cal L}$ is an isomorphism from the group $\tl{\cal L}$ onto $\Z^n\subset\R^n$.   
\label{rho-iso}
\end{lm}

\subsubsection{Relations between distinct tame subgroups of $\cal HG$ \label{hg}}


Let $\tl{\cal L}_a$ ($a=1,\ldots,m$) be a tame subgroup of the hoop group $\cal HG$ generated by independent loops ${\cal L}_a=\{l_{a,1},\ldots,l_{a,n_a}\}$. Applying a construction described in \cite{al-hoop} we can get a set ${\cal L}'=\{{l'}_1,\ldots,{l'}_n\}$ of independent loops  such that:
\begin{enumerate}
\item  every loop belonging to $\bigcup_{a=1}^m {\cal L}_a$ is a composition of loops belonging to $\tl{\cal L}'$,
\item given a loop $l\in\bigcup_{a=1}^m {\cal L}_a$ there exists a loop $l'_i\in{\cal L}'$ such that  
\begin{enumerate}
\item the loop $l$  can be decomposed as $l= k_1\circ{l'}^{\pm 1}_i\circ k_2$, where $k_1, k_2$ are loops built from the ones in ${\cal L}'$ {\em except} ${l'}_i$. Without loss of generality we will always assume that the  orientation of the loop  $l$ is such that $l= k_1\circ l'_i\circ k_2$.     
\item if the loop $l\in {\cal L}_a$ then it is the only loop in ${\cal L}_a$ in whose decomposition the loop ${l'_i}$ appears.
\end{enumerate}
\end{enumerate}
We caution the reader that throughout the paper we will use such decompositions only. 

Decompose each loop $l\in\bigcup_{a=1}^n {\cal L}_a$ in terms of the loops $\{{l'}_1,\ldots,{l'}_n\}$. Thus we obtain the following decomposition of hoops:  
\begin{equation}
\tilde{l}_{a,\nu}=\prod_{i=1}^{n}(\tilde{{l'}}_i)^{M^a_{\nu i}}:=(\tilde{{l'}}_1)^{M^a_{\nu 1}}\circ\ldots\circ(\tilde{{l'}}_n)^{M^a_{\nu n}},
\label{matrix}
\end{equation}
where every component of every matrix $M^a=(M^a_{\nu i})$ belongs to $\Z$. Next choose and fix an arbitrary set ${\cal L}_a$. By means of the map $\rho\equiv\rho_{\cal L'}:\tl{\cal L}'\rightarrow\R^n$ we get from (\ref{matrix}): 
\begin{equation}
\rho(\tilde{l}_{a,\nu})=\sum_{i=1}^n M^a_{\nu i}\,\varepsilon_i, \ \ {\text{hence}} \ \ (M^a_{\nu i})=(\rho(\tll_{a,\nu})_i).
\label{rho-M}
\end{equation}
Clearly, the matrix $M^a$ has  $n$ columns and $n_a$ rows. The properties of the decomposition of the loops $\{l_{a,1},\ldots,l_{a,n_a}\}$ in terms of $\{{l'}_1,\ldots,{l'}_n\}$ imply that properly ordered columns of the matrix  compose a unit $(n_a\times n_a)$-matrix. This means that the rank of $M^a$ is maximal and equal to $n_a$. Hence with each set ${\cal L}_a$ we can associate a linear subspace $V_a\subset\R^n$ whose basis is constituted by the vectors $\{\rho(\tll_{a,1}),\ldots,\rho(\tll_{a,n_a})\}$:
\begin{equation}
V_a:={\rm span}\{\rho(\tll_{a,1}),\ldots,\rho(\tll_{a,n_a})\}
\label{va}
\end{equation}

It turns out that the space $V_a$ describes the subgroup $\tl{\cal L}_a$ generated by ${\cal L}_a$ in an unambiguous way:
\begin{lm}
Let tame subgroups $\tl{\cal L}_a,\tl{\cal L}_{a'}$ of the hoop group $\cal HG$ and subspaces $V_a, V_{a'}$ of $\R^n$ be generated respectively by sets ${\cal L}_a$, ${\cal L}_{a'}$ of independent loops. Then  
\[
V_a=V_{a'}\Longleftrightarrow \tl{\cal L}_a=\tl{\cal L}_{a'}.  
\]
\label{l=va}
\end{lm}    
{\bf Proof.} Let us begin with  the implication $\Longrightarrow$. If $V_a=V_{a'}$ then $n_a=n_{a'}$ and there exists an invertible ($n_a\times n_a$)-matrix $Q_{\nu\mu}$ which transforms  the basis $\{\rho(\tll_{a,1}),\ldots,\rho(\tll_{a,n_0})\}$ onto the basis $\{\rho(\tll_{a',1}),\ldots,\rho(\tll_{a',n_0})\}$:
\[
\rho(\tll_{a',\mu})=\sum_{\nu=1}^{n_a}Q_{\mu\nu}\rho(\tll_{a,\nu}), \ \ \text{that is}\ \ M^{a'}_{\mu i}=\sum_{\nu=1}^{n_a}Q_{\mu\nu}M^{a}_{\nu i}
\] 
The facts that all components of the matrices $M^{a}$ and $M^{a'}$ are integers and that properly ordered columns of each of them compose a unit  $(n_a\times n_a)$-matrix imply  that the components of both matrices $Q_{\nu\mu}$ and $(Q^{-1})_{\nu\mu}$ are also integers. Hence by virtue of Lemma \ref{rho-iso} we have: 
\begin{equation}
\begin{gathered}
\rho(\tll_{a,\mu})=\sum_{\nu=1}^{n_a}Q_{\mu\nu}\rho(\tll_{a',\nu})\ \ \text{if and only if} \ \  \tll_{a,\mu}=\prod_{\nu=1}^{n_a}(\tll_{a',\nu})^{Q_{\mu\nu}},\\
\rho(\tll_{a',\mu})=\sum_{\nu=1}^{n_a}Q^{-1}_{\mu\nu}\rho(\tll_{a,\nu}) \ \ \text{if and only if} \ \ \tll_{a',\mu}=\prod_{\nu=1}^{n_a}(\tll_{a,\nu})^{Q^{-1}_{\mu\nu}}.
\end{gathered}
\label{Q}
\end{equation}
This means, that $\tl{\cal L}_a=\tl{\cal L}_{a'}$. 

To show the implication $\Longleftarrow$ it is enough to notice that $\tl{\cal L}_a=\tl{\cal L}_{a'}$ implies  \eqref{Q}. $\blacksquare$   

We have as well:
\begin{lm}
\[
\det Q=\pm 1
\]
\label{det-Q}
\end{lm} 
{\bf Proof.} The components $Q_{\nu\mu}$ and $Q^{-1}_{\nu\mu}$ are integers, hence $\det Q \in \Z$ and $\det (Q^{-1})=(\det Q)^{-1} \in \Z$. $\blacksquare$

\subsubsection{Diffeomorphism invariance of ${\cal HG}$ \label{hg-diff}}

Let us finally consider the dependence of the hoop group ${\cal HG}\equiv {\cal HG}_y$ on the choice of the base point $y\in \Sigma$ of the loops in $\mathbb{L}_y$. Clearly, this issue is equivalent to the one of diffeomorphism invariance of ${\cal HG}_y$. Consider then an analytic diffeomorphism $\phi$ of $\Sigma$ and suppose  that $\phi(y)=y'$. Let $l$ be an element of $\mathbb{L}_y$. Then $\phi(l)$ belongs to $\mathbb{L}_{y'}$. Evidently, the following map
\begin{equation}
{\cal HG}_y\ni \tl{l}\mapsto \tl{\phi}(\tl{l}):=\widetilde{\phi(l)}\in {\cal HG}_{y'}
\label{diff-hoop}
\end{equation}
is an isomorphism between the two hoop groups. However, as we are going to show now, every element of ${\cal HG}_{y'}$ defines an element of ${\cal HG}_{y}$ in a natural way. 

Assuming that $y\neq y'$ let us consider   piecewise analytic paths ${e}_1,{e}_2$ which originate at $y$ and end at $y'$. Then  the oriented loops
\[
l_1:={e}^{-1}_1\circ \phi(l)\circ {e}_1, \ \ l_2=:{e}^{-1}_2\circ \phi(l)\circ {e}_2, \ \ l':={e}_2^{-1}\circ{e}_1
\]     
are based at the point $y$. Since $l_2=l'\circ l_1\circ l^{\prime -1}$  and  the hoop group ${\cal HG}_y$ is commutative we have
\[
\tl{l}_2=\tl{l'}\circ\tl{l}_1\circ\tl{l'}^{-1}=\tl{l}_1 \in{\cal HG}_y .
\]   
Thus we see that the loop $\phi(l)$ based at the point $y'\neq y$ defines precisely one hoop in ${\cal HG}_y$. Therefore we are allowed to identify the hoop groups ${\cal HG}_y$ and ${\cal HG}_{y'}$ and treat the map \eqref{diff-hoop} as an automorphism of ${\cal HG}_y$. 
 
\subsection{Generalized $\R$-connections}
	
Every cylindrical function depends on holonomies along a {\em finite} set of hoops. Thus in this subsection we will consider sets of holonomies given by the tame subgroups of $\cal HG$ (recall that every tame subgroup of $\cal HG$ is generated by a {\em finite} number of hoops).  

Following \cite{al-hoop} we define\footnote{We emphasize that the construction of the Hilbert space does not require the generalized connections. We introduce them only by analogy to the case of a principal bundle with a compact structure group, where generalized connections appear naturally. However, if we decided to restrict ourselves to smooth connections then the notions considered in the present Subsection would be only slightly modified.}:
\begin{df}
Any group homomorphism from the hoop group $\cal HG$ into the group $\R$ is said to be a generalized connection on the bundle $P=\Sigma\times \R$. 
\end{df}
The space of all generalized connections on $P=\Sigma\times\R$ will be denoted by $\Abar$. The space $\Abar$ contains the space $\A$ of smooth connections on the bundle as its subspace. This is because the map
\[
{\cal HG}\ni \tll\mapsto A(\tll):=H(l,A)\in \R,
\]   
where $H(l,A)$ is the holonomy of the connection $A$ along the loop $l$, is a group homomorphism.

Given a tame subgroup $\tl{\cal L}$ of $\cal HG$ generated by a set  ${\cal L}=\{l_1,\ldots,l_n\}$ of independent loops we define the following equivalence relation on $\Abar$: we say that generalized connections  $\bar{A}_1,\bar{A}_2$ are equivalent,
\[
\bar{A}_1\sim\bar{A}_2 \ \ \ \text{if and only if} \ \ \ \bar{A}_1(\tll)=\bar{A}_2(\tll)
\]
for every hoop $\tll\in\tl{\cal L}$ \cite{al-hoop}. In what follows the quotient space $\Abar/\sim$ will be denoted by $\Abar/\tl{\cal L}$, and  $\pi_{\tl{\cal L}}$ will stand for the canonical projection from $\Abar$ onto $\Abar/\tl{\cal L}$, i.e. for the map $\bar{A}\mapsto [\bar{A}]$. We have the following lemma \cite{al-hoop} which is a direct consequence of Lemma \ref{x-Al}:
\begin{lm}
(i) A map  
\begin{equation}
\Abar/\tl{\cal L}\ni [\bar{A}]\mapsto {\cal I}_{\cal L}([\bar{A}]):=(\bar{A}(\tl{l}_1),\ldots,\bar{A}(\tl{l}_n))=(x_1,\ldots,x_n)\in\R^n.
\label{A-xx}
\end{equation}
is a bijection. (ii) Every equivalence class  $[\bar{A}]$ contains a smooth connection $A\in\A$. 
\label{AL-Rn}
\end{lm}

We emphasize that, given a quotient space $\Abar/\tl{\cal L}$, the map ${\cal I}_{\cal L}$ is not canonical since it depends on the choice of independent loops $\cal L$ generating the tame group $\tl{\cal L}$. Taking into account Equations \eqref{Q} we conclude that if two sets of independent loops $\cal L$ and $\cal L'$ generate the same tame subgroup $\tl{\cal L}$ then the corresponding maps ${\cal I}_{\cal L},{\cal I}_{\cal L'}$ are related to each other as follows
\begin{equation}
{\cal I}_{\cal L}([\bar{A}])=(\bar{A}(\tl{l}_i))=(\sum_{j=1}^n Q_{ij}\bar{A}(\tl{l}'_j))=(Q\circ{\cal I}_{\cal L'})([\bar{A}]).
\label{I-Q}
\end{equation}
As we see the change of a set of independent loops manifests by a {\em linear} invertible map $Q:\R^n\mapsto \R^n$ which is given by the matrix $(Q_{ij})$. Thus the linear structure on $\Abar/\tl{\cal L}$ is naturally defined.

Notice now that in our case the tame group $\tl{\cal L}\cong \Z^n$ can be embedded in $\Abar/\tl{\cal L}\cong\R^n$. Indeed, we  have
\[
\rho_{\cal L}(\tll_i)={\cal I}_{\cal L}([\bar{A}_i]),
\]
where $\{[\bar{A}_i]\}$ are elements of $\Abar/\tl{\cal L}$ such that $\bar{A}_i(\tll_j)=\delta_{ij}$. Thus the embedding is given by $\tll_i\mapsto [A_i]$. In the sequel we will often take advantage of this fact.
 
Let us now establish some relations between distinct spaces $\Abar/\tl{\cal L}$. Consider two sets ${\cal L}'=\{l'_1,\ldots,l'_{n'}\}$, ${\cal L}=\{l_1,\ldots,l_{n}\}$ of independent loops which generate tame groups $\tl{\cal L}'$ and $\tl{\cal L}$ respectively, such that $\tl{\cal L}'$ is a subgroup of $\tl{\cal L}$. Then   
\begin{equation}
\tilde{l}'_{\mu}=\prod_{i=1}^n (\tilde{{l}}_i)^{M_{\mu i}}, \ \ M_{\mu i}\in \Z.
\label{decompos}
\end{equation}
Acting on both sides of the equation with a connection $\bar{A}$ we get
\begin{equation}
\bar{A}(\tl{l}'_\mu)=\sum_{i=1}^n M_{\mu i}\bar{A}(\tl{l}_i),
\label{AMA}
\end{equation}
which can be expressed as  ${\cal I}_{\cal L'}= M\circ {\cal I}_{\cal L}$, where $M:\R^n\rightarrow\R^{n'}$ is a  linear map given by the matrix $(M_{\mu i})$. According to Lemma \ref{AL-Rn} the map ${\cal I}_{\cal L'}$ is a bijection, hence the map $M$ is surjective.  Therefore 
\begin{equation}
\pi_{\tl{\cal L}'\tl{\cal L}}:={\cal I}^{-1}_{\cal L'}\circ M\circ {\cal I}_{\cal L}
\label{IMI}
\end{equation}
is a map\footnote{The map $\pi_{\tl{\cal L}'\tl{\cal L}}$ does not depend on the choice of the sets ${\cal L}'$ and ${\cal L}$ of independent loops generating the tame groups $\tl{\cal L}'$ and $\tl{\cal L}$. This is because the change of the sets ${\cal L}'$ and ${\cal L}$ manifests by an appropriate change of the map $M$ which follows from \eqref{decompos}.} from $\Abar/{\cal L}$ {\em onto} $\Abar/{\cal L'}$ such that
\begin{equation}
\pi_{\tl{\cal L}'\tl{\cal L}}\circ \pi_{\tl{\cal L}}=\pi_{\tl{\cal L}'}.
\label{pi_LL}
\end{equation}

\subsection{Schwarz functions} 

Recall that we are going to build the Hilbert space by means of cylindrical functions given by functions square-integrable with respect to the Lebesgue measure $dx$ on $\R^n$. The space $L^2(\R^n,dx)$ contains as a dense subset the space of Schwarz functions whose properties (described by Lemma \ref{3-stat} and Theorem \ref{lnzfunk}) will be essential  while constructing the Hilbert space.  

Let $\bld{\a}$ denote a multilabel $(\a_1,\ldots,\a_n)$ such that every $\alpha_i$ belongs to $\N_0:=\{0,1,2,\ldots\}$. Given a smooth function $f:\R^n\rightarrow\C$, we denote by $D^{\bld{\a}}$ a partial derivative
\[
D^{\bld{\a}}f:=\frac{\partial^{\sum_i \a_i}}{\partial x^{\a_1}_1\ldots \partial x^{\a_n}_n} f,
\] 
where $(x_i)$ are the Cartesian coordinates on $\R^n$.  
\begin{df}
A map $f:\R^n\rightarrow\C$ is said to be  a Schwarz function if and only if  $f\in C^{\infty}(\R^n)$ and for every $m\in \N_0$ and for every derivative $D^{\bld{\a}}f$ 
\begin{equation}
\lim_{r\rightarrow\infty} {(D^{\bld{\a}}f)}\,{r^{m}}=0,
\label{granica}
\end{equation}
where $r=\sqrt{\sum_i x^2_i}$. 
\end{df}

The set of Schwarz functions on $\R^n$, denoted by $\S(\R^n)$,  possesses a natural structure of a $*$-algebra.

It is easy to verify the following lemma:
\begin{lm}
(i) Suppose that $B$ is a linear automorphism of $\R^n$. Then a map
\[
\S(\R^n)\ni f \mapsto B^* f:=f\circ B \in \S(\R^n)
\]
is an automorphism of the  $*$-algebra  $\S(\R^n)$. (ii) Suppose that $f$ is a non-zero element of $\S(\R^n)$. For any set $\{X_1,\ldots, X_m\}$ of non-zero constant vector fields on $\R^n$ the derivative
\[
(X_1\ldots X_m)f\neq 0.
\] 
\label{der-lm}
\end{lm}

Let us fix  a {\em surjective} linear map $M$ from $\R^n$ onto $\R^{n'}$ $(n'\leq n)$. We define a set $\S_M(\R^n)$ of maps given by a pull-back of the ones in $\S(\R^{n'})$:
\[
\S_M(\R^n):=\{\ f:\R^n\mapsto \C \ | \ f=M^* f', \ f'\in \S(\R^{n'})\ \}.  
\]

Given $f\in\S_M(\R^n)$, denote by $\tilde{V}(f)$ a linear subspace of $\R^n$ such that every element $x$ of $\tilde{V}(f)$ defines a constant vector field $X$ (i.e. $X(0)=x$) on $\R^n$ satisfying
\begin{equation}
Xf=0.
\label{Xf-0}
\end{equation} 
In fact all non-zero functions $f\in \S_M(\R^n)$ define the same subspace of $\R^n$. To see this consider the canonical bases:  $(\varepsilon_i)$ of $\R^n$ and $(\varepsilon'_{\mu})$  of $\R^{n'}$. Let
\[
M(\varepsilon_i)=\sum_{\mu=1}^{n'} M_{\mu i}\varepsilon'_\mu.
\] 
Then Equation \eqref{Xf-0} can be written as follows
\[
Xf=X(M^*f')=M^* [ (M_*X) f']=0, \ \ \text{hence} \ \ (M_*X) f'=0.
\]
Clearly, $M_*X$ is a constant vector field on $\R^{n'}$. Now the statement $(ii)$ of Lemma \ref{der-lm} applied to the non-zero Schwarz function $f'$ implies that $M_*X=0$ or, equivalently, that $\sum_{i=1}^n M_{\mu i} X_i=0$. The last expression gives us, in particular,
\[
\sum_{i=1}^n M_{\mu i} X_i(0)=\sum_{i=1}^n M_{\mu i} x_i=0
\]
and, consequently, 
\begin{equation}
\tilde{V}(f)=\ker M.
\label{Vf-ker}
\end{equation}
Now we can prove:
\begin{lm}
The following statements are equivalent:
\begin{gather}
\S_{M_1}(\R^n)\cap\S_{M_2}(\R^n)\neq\{0\}\tag{$i$}\\
\ker M_1=\ker M_2\tag{$ii$}\\
\S_{M_1}(\R^n)=\S_{M_2}(\R^n)\tag{$iii$}
\end{gather}
\label{3-stat}
\end{lm}
{\bf Proof.} Let us show first that $(i)$ implies $(ii)$. Consider a non-zero $f\in\S_{M_1}(\R^n)\cap\S_{M_2}(\R^n)$. By virtue of \eqref{Vf-ker}  $\tilde{V}(f)=\ker M_1$ and $\tilde{V}(f)=\ker M_2$. To show that $(iii)$ follows $(ii)$ notice that if $\ker M_1=\ker M_2$ then ${\rm im}\,M_1={\rm im}\, M_2=\R^{n'}$ and there exists a linear automorphism $Q$ of $\R^{n'}$ such that $Q\circ M_1=M_2$. Now the statement $(i)$ of Lemma \ref{der-lm} implies $(iii)$ immediately. $\blacksquare$  

The lemma just proved means that the spaces $\S_M(\R^n)$ can unambiguously be labelled  by linear subspaces of $\R^n$, therefore we will write $\S_{\tilde{V}}(\R^n)$ instead of $\S_M(\R^n)$, provided $\ker M=\tl{V}$.  

We have the following:
\begin{thr}
Suppose that $\{\tl{V}_1,\ldots,\tilde{V}_m\}$  is a set of subspaces of $\R^n$ such that $\tilde{V}_a=\tilde{V}_b$ if and only if $a=b$. Then non-zero functions $\{f_1,\ldots,f_m\}$ such that $f_a\in \S_{\tilde{V}_a}(\R^n)$ are linearly independent.
\label{lnzfunk}
\end{thr}
{\bf Proof.} Denote by ${\cal E}_a=\{\varepsilon_{a,i}\}$ a finite set of vectors spanning the space $\tilde{V}_a$ (${\cal E}_a$ does not have to be a basis of the space). Assume that 
\begin{enumerate}
\item the sets $\{{\cal E}_a\}$ are chosen in such a way\footnote{It can be achieved as follows: let $\tl{{\cal E}}_a=\{\tl{\varepsilon}_{a,i}\}$ be a basis of $\tl{V}_a$; then set ${\cal E}_a:=\tl{V}_a\cap (\bigcup_b \tl{{\cal E}}_b)$.} that if $\varepsilon_{a,i}\in\tilde{V}_{a'}$, then $\varepsilon_{a,i}\in {\cal E}_{a'}$.
\item the spaces $\{\tl{V}_1,\ldots,\tilde{V}_m\}$ are labelled in such a manner that $\dim \tilde{V}_a\leq \dim\tilde{V}_{a+1}$.
\end{enumerate}

Every vector $\varepsilon_{a,i}$ belonging to  ${\cal E}:=\bigcup_{a=1}^k {\cal E}_a$ defines a constant vector field $X_{a,i}$ on $\R^n$. Let us define an operator acting on  smooth functions on $\R^n$:
\[
\hat{{\cal E}}_{a_0}f:= (X_{b,i}\ldots X_{b',i'})f,
\]
where the constant vector fields $\{X_{b,i},\ldots, X_{b',i'}\}$ on $\R^n$ are defined by vectors in ${\cal E}\setminus {\cal E}_{a_0}$.

The theorem we are proving now is a consequence of the following fact:
\begin{equation}
\hat{{\cal E}}_{a_0}f_a\neq0\Longleftrightarrow \tilde{V}_a\subset\tilde{V}_{a_0}.
\label{E-V}
\end{equation}
Indeed, to show the implication $\Longrightarrow$ assume that $\hat{{\cal E}}_{a_0}f_a\neq0$ and $\tilde{V}_a\not\subset\tilde{V}_{a_0}$. Then there exists a vector $\varepsilon_{a,i}$ which  belongs to ${\cal E}_a$ and does not belong to ${\cal E}_{a_0}$, hence $\varepsilon_{a,i}\in {\cal E}\setminus {\cal E}_{a_0}$. Then $f_a$ is annihilated by $\hat{{\cal E}}_{a_0}$. 

To show the implication $\Longleftarrow$ of \eqref{E-V} recall that, given $a$, there exists a surjective linear map $M_a:\R^n\mapsto \R^{n_a}$ and a map $f'_a\in \S(\R^{n_a})$ such that 
\[
\ker M_a=\tl{V}_a \ \ \text{and} \ \ f_a=M_a^* f'_a.
\]
Notice now that the assumption 1 implies that the vectors in ${\cal E}\setminus {\cal E}_{a_0}$ do not belong to $\ker M_a$. Therefore the constant vector fields on $\R^n$ defined  by the vectors can be pushed forward by $M_a$ to some {\em non-zero} constant vector fields on $\R^{n_a}$. Thus we obtain
\[
\hat{{\cal E}}_{a_0}f_a=(X_{b,i}\ldots X_{b',i'})M^*_a f'_a=M^*_a [\ (M_a)_*X_{b,i}\ldots (M_a)_*X_{b',i'}\ ] f'_a
\] 
By virtue of the statement $(ii)$ of Lemma \ref{der-lm} the r.h.s. of the above equation is non-zero and the implication follows.


Let
\[
\sum_{a=1}^m\lambda_a f_a=0,
\]
where $\{\lambda_a\}$ are complex numbers. Acting  on both sides of the above  equation by the operators $\{\hat{{\cal E}}_a\}$ in turn determined by the assumption 2 and using \eqref{E-V} we get $\lambda_a=0$ for all $a=1,\ldots,m$. $\blacksquare$

\subsection{Cylindrical Schwarz function \label{cyl-schw}}

Let us introduce now the notion of cylindrical Schwarz functions, from which the Hilbert space over the space of $\R$-connection will be built. 

\begin{df}
Let a tame subgroup $\tl{\cal L}$ of $ {\cal HG}$ be generated by a set  ${\cal L}=\{l_1,\ldots,l_n\}$ of independent loops. A map $\Psi:\Abar\rightarrow\C$ is said to be a smooth cylindrical function compatible with the group $\tl{\cal L}$ if and only if  
\begin{equation}
\Psi=(\pi^\ast_{\tl{\cal L}}\circ{\cal I}^\ast_{\cal L})\psi,
\label{fun-cyl}
\end{equation}
where $\psi\in C^{\infty}(\R^n)$, i.e.:
\[
\Psi(\bar{A})=\psi(\bar{A}(l_1),\ldots,\bar{A}(l_n)).
\]     
A cylindrical functions $\Psi$ given by \eqref{fun-cyl}, where $\psi\in \S(\R^n)$, is said to be a cylindrical Schwarz function compatible with $\tl{\cal L}$.
\end{df}
The linear space of all cylindrical (cylindrical Schwarz) functions  compatible with the tame group $\tl{\cal L}$  will be denoted by $\cyl(\Abar/\tl{\cal L})$ ($\S(\Abar/\tl{\cal L})$). Clearly, by virtue of \eqref{I-Q} the spaces $\cyl(\Abar/\tl{\cal L})$ and $\S(\Abar/\tl{\cal L})$ do not depend on the choice of the bijection $\cal I_{\cal L}$ in \eqref{fun-cyl}. 

There immediately arises a natural question: can a cylindrical Schwarz function be compatible with two distinct tame groups $\tl{\cal L}_1$ and $\tl{\cal L}_2$? The answer if provided by the following 

\begin{thr}
Suppose that a  non-zero map $\Psi:\Abar\rightarrow \C$ is a cylindrical Schwarz function compatible with  $\tl{\cal L}_1$ and $\tl{\cal L}_2$. Then $\tl{\cal L}_1=\tl{\cal L}_2$.     
\label{uniq}
\end{thr}
{\bf Proof.} The assumption of the theorem means that there exist maps $\psi_a:\R^{n_a}\mapsto \C$ ($a=1,2$) such that 
\[
\Psi=(\pi^*_{\tl{\cal L}_1}\circ {\cal I}^*_{{\cal L}_1})\psi_1=(\pi^*_{\tl{\cal L}_2}\circ {\cal I}^*_{{\cal L}_2})\psi_2,
\]
where the set ${\cal L}_a=\{l_{a, 1},\ldots,l_{a, n_a}\}$  of independent loops generates the tame group $\tl{\cal L}_a$. There exists a set ${\cal L}=\{{l}_1,\ldots,{l}_n\}$ of independent loops such that the loops in ${\cal L}_1$ and ${\cal L}_2$ can be decomposed into the loops in ${\cal L}$ in the way described in Subsection \ref{hg}. Then using Equations \eqref{IMI} and \eqref{pi_LL} and surjectivity of the projection $\pi_{\tl{\cal L}}$ we get from the above expression
\[
(M^1)^*\psi_1=(M^2)^*\psi_2=:\psi
\]
where $M^a$ is a map from $\R^n$ {\em onto} $\R^{n_a}$ given by the decomposition of loops in ${\cal L}_a$ in terms of the ones in $\cal L$. Thus $\psi\in\S_{M^a}(\R^{n})$ and by virtue of Lemma \ref{3-stat}  
\[
\ker M^1=\ker M^2 \ \ \text{and} \ \ n_1=n_2.
\]
Thus there exists a linear automorphism $Q$ of $\R^{n_a}$ such that
\[
Q\circ M^1=M^2.
\]  

Consider now the map $\rho_{\cal L}:\tl{\cal L}\rightarrow \R^n$ given by \eqref{ln}. Using \eqref{rho-M} we can easily express the last equation in the form \eqref{Q} as it was done in the proof of Lemma \ref{l=va}. Thus the theorem follows. $\blacksquare$  

We also have:
 
\begin{thr}
Suppose that $\tl{\cal L}_a$ ($a=1,\ldots,m$) are tame subgroups of $\cal HG$ such that  $\tl{\cal L}_a= \tl{\cal L}_{a'}$ if and only if  $a = a'$. Then non-zero functions $\{\Psi_a\}$ such that $\Psi_a\in \S(\Abar/\tl{\cal L}_a)$ are linearly independent.  
\label{lin-ind}
\end{thr}
{\bf Proof.} Let ${\cal L}_a$ be a set of independent loops generating the tame group $\tl{\cal L}_a$. Decompose the loops in $\bigcup_a {\cal L}_a$ in terms of independent loops ${\cal L}=\{l_1,\ldots,l_n\}$ as described in Subsection \ref{hg}.

Every function $\Psi_a$ is defined by a function $\psi_a\in\S(\R^{n_a})$ according to \eqref{fun-cyl}. On the other hand the map $\rho\equiv\rho_{{\cal L}}$ given by \eqref{ln} defines a surjective map $M^a:\R^n\mapsto \R^{n_a}$ according to  \eqref{rho-M}. Therefore
\[
\psi'_a:=(M^{a})^*\psi_a\in \S_{\tl{V}_a}(\R^n),
\]    
where $\tl{V}_a=\ker M^a$. Thus a vector $x=(x_i)$ in $\R^n$ belongs to $\tl{V}_a$ if and only if 
\[
0=\sum_{i=1}^n M^a_{\mu i} x_i = \sum_{i=1}^n \rho(\tll_{a,v})_i x_i=\scal{\rho(\tll_{a,v})}{x},
\]   
where the scalar product is the canonical one on $\R^n$. Therefore we have
\begin{equation}
\ker M^a=\tl{V}_a=V^\perp_a,
\label{perp-V}
\end{equation}
where $V_a$ is spanned by the vectors $\{\rho(\tll_{a,v})\}$ (see \eqref{va}).  

Now Lemma \ref{l=va} and Theorem \ref{lnzfunk} immediately imply that the functions $\{\psi'_a\}$ are linearly independent and consequently $\{\Psi_a\}$ are so. $\blacksquare$

\subsection{Space of cylindrical Schwarz functions}

Let $\h_0$ be a space of complex functions on $\Abar$ of the form:
\[
\Psi=\sum_{a=1}^n \Psi_a,   
\]
where $\Psi_a\in\S(\Abar/\tl{\cal L}_a)$ and $\tl{\cal L}_a$ is a tame subgroup of ${\cal HG}$. By Theorem \ref{lin-ind} we can write:
\begin{equation}
\h_0=\bigoplus_{\tl{\cal L}\subset{\cal HG}}\S(\Abar/\tl{\cal L}).
\label{dec-h0}
\end{equation}
In the next section we will define a scalar product on $\h_0$ and after an appropriate completion we will obtain a Hilbert space $\h$ over $\Abar$.

It is clear that $\h_0$ is a linear space. However, it is not an algebra, at least not under the multiplication
\[
(\Psi\Psi')(\bar{A})=\Psi(\bar{A})\Psi'(\bar{A}).
\]
In Appendix \ref{h0-notal} we give an example of two functions belonging to $\h_0$ such that the product of them is not an element of the space.

\section{Construction of the Hilbert space}

Theorems \ref{uniq} and \ref{lin-ind} simplify the task of constructing a Hilbert space over cylindrical Schwarz functions on $\Abar$. Notice that since every such  a function is compatible  with precisely one tame subgroup $\tl{\cal L}$ of $\cal HG$, a scalar product of two functions compatible with $\tl{\cal L}$ can be defined in the manner described in the introduction, that is, by a measure on $\Abar/\tl{\cal L}$. On the other hand  cylindrical Schwarz functions compatible with distinct tame subgroups are linearly independent. Therefore if $\tl{\cal L}_1\neq\tl{\cal L}_2\neq\tl{\cal L}_3$ then the scalar product of elements of $\S(\Abar/\tl{\cal L}_1)$ and $\S(\Abar/\tl{\cal L}_2)$  can be defined independently of the scalar product between e.g. functions in $\S(\Abar/\tl{\cal L}_1)$ and $\S(\Abar/\tl{\cal L}_3)$ etc.

Now, following the above discussion, we are going to define a scalar product on the space $\h_0$ introduced at the end of the previous Subsection. 

\subsection{Haar measure on $\Abar/\tl{\cal L}$}

Lemma \ref{AL-Rn} and the discussion below it ensure that there is a natural Lie group structure on $\Abar/\tl{\cal L}$ isomorphic to $\R^n$, where $n$ is the number of the generators of the tame group $\tl{\cal L}$. Therefore it is natural to use a Haar measure on $\Abar/\tl{\cal L}$ to define a scalar product on $\S(\Abar/\tl{\cal L})$. Notice that the group structure on $\Abar/\tl{\cal L}$ is defined by means of the map ${\cal I}_{\cal L}$ \eqref{A-xx} which depends on the choice of a set of independent loops generating the tame group $\tl{\cal L}$. Consequently, any Haar measure $d\mu_{\tl{\cal L}}$ on $\Abar/\tl{\cal L}$ is determined as
\begin{equation}
\int_{\Abar/\tl{\cal L}}\Psi\, d\mu_{\tl{\cal L}} := \int_{\R^n} [\ ({\cal I}^{-1}_{\cal L})^* \Psi\ ] \ c\,dx,
\label{haar-R}
\end{equation}
where $c>0$ is an arbitrary constant, $dx=dx_1\ldots dx_n$ is the Lebesgue measure on $\R^{n}$, and $(x_1,\ldots,x_n)$ are the Cartesian coordinates on $\R^n$ given by \eqref{A-xx}.   

The above expression suggests that, given tame group $\tl{\cal L}$, the measure $d\mu_{\tl{\cal L}}$ might depend on the choice of the set ${\cal L}$.  To show that it is not the case consider two sets $\cal L$, $\cal L'$ generating the group $\tl{\cal L}$ and require
\[
 \int_{\R^n} [\ ({\cal I}^{-1}_{{\cal L'}})^* \Psi\ ] \ c\,dx'= \int_{\R^n} [\ ({\cal I}^{-1}_{{\cal L}})^* \Psi\ ] \ c\,dx
\]      
for any $\Psi\in\S(\Abar/\tl{\cal L})$. Using Equation \eqref{I-Q} we obtain 
\[
 \int_{\R^n} [\ Q^*\circ({\cal I}^{-1}_{{\cal L}})^* \Psi\ ] \ c\,dx'= \int_{\R^n} [\  ({\cal I}^{-1}_{{\cal L}})^* \Psi\ ] \ c\,dx.
\]
Taking into account the second equation of \eqref{Q} and Equation \eqref{A-xx} we conclude that the linear operator $Q$ defines a linear transformation of the coordinates $(x'_1,\ldots,x'_n)$ onto $(x_1,\ldots,x_n)$. Now, Lemma \ref{det-Q} guarantees that the above equation holds and, consequently, that the Haar measure $d\mu_{\tl{\cal L}}$ on $\Abar/\tl{\cal L}$ is unambiguously defined.  

\subsection{Scalar product on $\h_0$}

As we have already noticed the decomposition \eqref{dec-h0} of the linear space $\h_0$ of all cylindrical Schwarz functions on $\Abar$ implies that in order to define a  scalar product $\scal{\cdot}{\cdot}$ on $\h_0$ it is necessary and sufficient to define a set of maps labelled by tame subgroups $\tl{\cal L},\tl{\cal L'}$ of $\cal HG$
\[
\scal{\cdot}{\cdot}_{\tl{\cal L}\tl{\cal L'}}\ : \ \S(\Abar/\tl{\cal L})\times\S(\Abar/\tl{\cal L'})\rightarrow \C,
\]     
such that each of them is antilinear in the first argument, and linear in the second one. Obviously, we will have to assure that the scalar product $\scal{\cdot}{\cdot}$ is $(i)$ positive definite and $(ii)$  diffeomorphism invariant. As we will show below, such a scalar product can be provided by the strategy described in Subsection \ref{outline}.


Thus, given $\Psi,\Psi'\in \S(\Abar/\tl{\cal L})$ and a Haar measure $d\mu_{\tl{\cal L}}$ on $\Abar/\tl{\cal L}$, we define
\begin{equation}
\scal{\Psi}{\Psi'}=\scal{\Psi}{\Psi'}_{\tl{\cal L}\tl{\cal L}}:=\int_{\Abar/\tl{\cal L}}\overline{\Psi}\Psi'd\mu_{\tl{\cal L}},
\label{integ}
\end{equation}
and for every pair $\tl{\cal L}_1,\tl{\cal L}_2$ such that $\tl{\cal L}_1\neq\tl{\cal L}_2$ we set
\begin{equation}
\scal{\cdot}{\cdot}_{\tl{\cal L}_1\tl{\cal L}_2}=0.
\label{zero2}
\end{equation}
We emphasize that because of the non-compactness of $\Abar/\tl{\cal L}$ there is no canonical choice (normalization) of the Haar measure and therefore the above scalar product cannot be defined canonically either. 

In this way we obtained a family of positive definite scalar products on $\h_0$ which differ from each other by the choice of Haar measures $d\mu_{\tl{\cal L}}$ assigned to every tame group $\cal L$. Given a member of the family, Theorem \ref{thr-zero} implies that the Hilbert space $\h$ obtained by the completion of $\h_0$ with respect to the norm $||\Psi||:=\sqrt{\scal{\Psi}{\Psi}}$ is an orthogonal sum:
\begin{equation}
\h=\bigoplus_{\tl{\cal L}\subset{\cal HG}}L^2(\Abar/\tl{\cal L},d\mu_{\tl{\cal L}}).
\label{ort-dec}
\end{equation}


\subsubsection{Flux operators on $\h_0$}

The choice of the map $\scal{\cdot}{\cdot}_{\tl{\cal L}_1\tl{\cal L}_2}=0$ for $\tl{\cal L}_1\neq\tl{\cal L}_2$ is the simplest one which gives a Hilbert space of the desired properties. However, this choice can be justify in another way. Let us recall that we are going to apply the Hilbert space \eqref{ort-dec} as the carrier space of a $*$-representation of cylindrical and flux functions. We are going to show now that the Hilbert space \eqref{ort-dec} is the only one (in the class of Hilbert spaces defined by those scalar products on $\h_0$ which satisfy \eqref{integ}) which admits a natural $*$-representation of flux functions.    

The phase space of any diffeomorphism invariant theory of $\R$-connections  consists of pairs $(A_\mu(y),$ $ \tilde{E}^\nu(y'))$ ($y,y'\in\Sigma$) of conjugated variables, where $A_\mu(y)\,dy^\mu$ is a differential one-form on $\Sigma$ corresponding to a smooth $\R$-connection $A$ on the bundle $\Sigma\times \R$, and $\tilde{E}^\nu(y')$ is a vector density. Following \cite{cq-diff} we define the only non-vanishing Poisson brackets of the variables as\footnote{One considers also other Poisson brackets of the variables --- see e.g. \cite{alex}.}
\[
\{A_\mu(y),\tl{E}^\nu(y')\}=\delta(y,y')\delta^\nu_\mu.
\]      
Taking advantage of the Levi-Civita anti-symmetric density $\epsilon_{\mu_1\ldots\mu_d}$ $(d=\dim\Sigma)$ we can integrate the momentum variable $\tl{E}^\nu$ over a $(d-1)$-di\-men\-sio\-nal {\em oriented analytic} surface $S$ embedded in $\Sigma$ 
\begin{equation}
E_S:=\int_S \tl{E}^\mu\ \epsilon_{\mu\mu_1\ldots\mu_{(d-1)}}\ dx^{\mu_1}\wedge \ldots\wedge dx^{\mu_{(d-1)}}\in\R,
\label{flux-f}
\end{equation}
obtaining as a result the flux (function) $E_S$ of the field $\tl{E}^\nu$ across the surface $S$. The regularization procedure described in \cite{area} provides us with a so called flux operator $\hat{E}_S$ as a natural counterpart of the flux function $E_S$. Assume now that $\Psi\in\cyl(\Abar/\tilde{\cal L})$, where $\tilde{\cal L}$ is generated by ${\cal L}=\{l_1,\ldots,l_n\}$, is given by a function $\psi\in C^{\infty}(\R^n)$ according to \eqref{fun-cyl}. Then the flux operators $\hat{E}_S$ acts on the function as follows:  
\begin{equation}
\hat{E}_S\Psi=-\frac{i}{2}(\pi^\ast_{\tl{\cal L}}\circ{\cal I}^\ast_{\cal L})\ (\sum_{j=1}^n n_j\frac{\partial}{\partial x_j} \psi).
\label{flux-op}
\end{equation}
The integers $\{n_j\}$ occurring in the above formula contain the information about the intersections of the loops $\{l_1,\ldots,l_n\}$ with the surface $S$. Given a loop $l_j$, we subdivide it on a finite number of (connected and oriented\footnote{The orientation of the segment is inherited from the orientation of the loop.}) segments such that each segment is either $(i)$ contained in $S$ (modulo its endpoints) or $(ii)$ the intersection of the segment with $S$ coincides with precisely one endpoint of the segment or $(iii)$ the segment does not intersect $S$. Let $n_j^+$ be the number of segments of the kind $(ii)$ which either are 'outgoing' from $S$ and placed 'above' the surface or are 'incoming' to $S$ and are placed 'below' the surface. Similarly, let $n_j^-$ be the number of segments of the kind $(ii)$ which either are 'outgoing' from $S$ and are placed 'below' the surface or are 'incoming' to $S$ and are placed 'above' the surface. Then
\[
n_j:=n_j^+-n_j^-.
\]   

It is clear that every flux operator preserves the spaces $\cyl(\Abar/\tl{\cal L})$ and $\S(\Abar/\tl{\cal L})$, hence it also preserves the space $\h_0$. Because every flux function \eqref{flux-f} is real it is natural to require a scalar product on $\h_0$ to guarantee symmetricity of every flux operator \eqref{flux-op}. 
  


\begin{thr}
Suppose that there exists a scalar product $\scal{\cdot}{\cdot}$ on $\h_0$ such that $\scal{\cdot}{\cdot}_{\tl{\cal L}\tl{\cal L}}$ is given by \eqref{integ} and  every flux operator $\hat{E}_S$ is symmetric with respect to the scalar product. Then for every pair $\tl{\cal L}_1,\tl{\cal L}_2$ of tame groups such that $\tl{\cal L}_1\neq \tl{\cal L}_2$
\begin{equation}
\scal{\cdot}{\cdot}_{\tl{\cal L}_1\tl{\cal L}_2}=0.
\label{zero3}
\end{equation}
\label{thr-zero}
\end{thr}
The proof of the theorem is quite technical and therefore it is relegated to Appendix \ref{thr-zero-ap}.

\subsubsection{Diffeomorphism invariance of the scalar product}

As we said in the introduction we require the scalar product on the Hilbert space $\h$ to be diffeomorphism invariant. In Subsection \ref{hg-diff} we defined a natural action of diffeomorphisms of $\Sigma$ on the hoop group $\cal HG$ which can be lifted to an action of diffeomorphisms on $\h_0$. Given tame subgroup $\tl{\cal L}$ of $\cal HG$ we define  
\begin{equation}
\S(\Abar/\tl{\cal L})\ni \ \Psi \ \mapsto \ \bG_{\phi} \Psi:=(\pi_{\tl{\phi}(\tl{\cal L})}^*\circ \,{\cal I}_{\phi({\cal L})}^*) \psi  \ \in \S(\Abar/\tl{\phi}(\tl{\cal L})),
\label{diff-cyl}
\end{equation}   
where $\Psi$ is given by \eqref{fun-cyl}, $\phi$ is an analytic diffeomorphism of $\Sigma$ and $\tl{\phi}$ is defined by \eqref{diff-hoop}.

It is clear that $\tl{\phi}$ maps two distinct tame groups onto two distinct ones. Hence $\bG_{\phi}$ preserves the scalar product $\scal{\cdot}{\cdot}_{\tl{\cal L}_1\tl{\cal L}_2}$ defined for every pair $\tl{\cal L}_1\neq \tl{\cal L}_2$ of tame groups. 

On the other hand the action $\bG_{\phi}$ preserves the scalar product $\scal{\cdot}{\cdot}_{\tl{\cal L}\tl{\cal L}}$ if and only if it maps the Haar measure on $\Abar/\tl{\cal L}$ defining the product (Equation \eqref{integ}) onto the appropriate one on $\Abar/\tl{\phi}(\tl{\cal L})$. Recall that the Haar measure on $\Abar/\tl{\cal L}$ is specified by the constant $c\equiv c_{\tl{\cal L}}$ occurring in the r.h.s. of  \eqref{haar-R}. It is easy to see that $\bG_{\phi}$ preserves the scalar product if and only if $c_{\tl{\cal L}}=c_{\tl{\phi}(\tl{\cal L})}$.

We conclude with the observation that the set of diffeomorphisms invariant scalar products  is labelled by sets of constants $\{c_{[\tl{\cal L}]}\}$, where $[\tl{\cal L}]$ denotes a class of tame subgroups obtained from $\tl{\cal L}$ by the action of all analytic diffeomorphisms of $\Sigma$. 

In this way we obtained a family of diffeomorphism invariant Hilbert spaces $\{\h\}$ built over the set of $\R$-connections and labelled by the sets of constants $\{c_{[\tl{\cal L}]}\}$. However, we can identify\footnote{This fact was pointed out to the author by Professor Jerzy Kijowski.} each pair of the Hilbert spaces by means of a natural unitary map. Assuming that a Hilbert space $\h$ is defined by the constants $\{c_{[\tl{\cal L}]}\}$, and $\h'$ --- by $\{c'_{[\tl{\cal L}]}\}$ we obtain the unitary map from $\h$ onto $\h'$ as the unique closure of
\[
\h_0\supset S(\Abar/\tilde{\cal L})\ni \ \psi \ \mapsto \ \sqrt{\frac{c_{[\tl{\cal L}]}}{c'_{[\tl{\cal L}]}}}\, \psi \ \in  S(\Abar/\tilde{\cal L})\subset \h_0.
\]

\subsection{Other diffeomorphism invariant scalar products on $\h_0$}

In the previous subsection we constructed a family of diffeomorphism invariant scalar products on $\h_0$ such that every two  Schwarz cylindrical functions compatible with {\em distinct} tame groups of hoops are mutually orthogonal. It is natural to ask now whether there exist diffeomorphism invariant, positive definite scalar products on $\h_0$ such that for some $\tl{\cal L}_1\neq \tl{\cal L}_2$   
\[
\scal{\cdot}{\cdot}_{\tl{\cal L}_1\tl{\cal L}_2}\neq 0.
\]
It turns out that such scalar products do exist which is proved in Appendix \ref{non-orth}.

\section{Discussion \label{disc}}

Let us remind  that our goal is not merely a diffeomorphism invariant Hilbert space  built from functions of connections with a non-compact structure group but such a Hilbert space equipped with a $*$-representation of the elementary variables. 
We have already found a $*$-representation of flux functions on the Hilbert space $\h$ (the representation is given by \eqref{flux-op}).  What remains is to choose a set of some cylindrical functions on $\Abar$ as elementary variables and find a $*$-representation of the functions. Unfortunately, {\em in practice} our possibilities to construct a representation of any cylindrical functions on the Hilbert space $\h$ are limited to the representation defined by the multiplication map: 
\begin{equation}
\hat{\Phi}\Psi= {\Phi}\Psi.
\label{repr-mult}
\end{equation}
The reason for that is threefold: firstly it is not clear how to define any other non-trivial representation of cylindrical functions on $\h$; secondly, any admissible representation of cylindrical functions together with the representation \eqref{flux-op} of flux functions have to define a representation of the elementary variables which satisfies the second of the conditions \eqref{star-rep}; thirdly, the resulting representation of the variables has to be diffeomorphism invariant\footnote{For the definition of diffeomorphism invariance of the representation see e.g. \cite{diff-inv}.}. Evidently, the representations \eqref{repr-mult} and \eqref{flux-op} meet these requirements.

It is easy to see that cylindrical functions playing the role of elementary variables have to be elements of $L^2(\Abar/\tl{\cal L},d\mu_{\tl{\cal L}})$ --- if e.g. $\Phi\in\cyl(\Abar/\tl{\cal L}_1)$ and $\Psi\in\S(\Abar/\tl{\cal L}_2)$ then in general $\Phi\Psi\not\in\h$. Thus a natural choice of elementary variables  is (except the flux functions) an algebra of cylindrical functions contained in the Hilbert space $\h$ as its (dense) subspace. However, the conclusion  that $\h_0$ is not an algebra (see Appendix \ref{h0-notal}) indicates towards some problems concerning the issue. Since multiplication by  Schwarz functions (including those of compact support) does not preserve the space $\h_0$ it is possible that it also does not preserve the Hilbert space $\h$. Thus we should be more specific in determining cylindrical functions as elementary variables. But Schwarz functions are quite specific and it is not clear, to which class of functions we should restrict ourselves. Thus the issue of the choice of those elementary variables which are defined on the configuration space (the space of connections) is left open.  

On the other hand one can hope that it is possible to extend the scalar product $\scal{\cdot}{\cdot}$ defined on $\h_0$ to the one on an algebra $\check{\h}_0$ generated by functions in $\h_0$. Then after an appropriate completion  we would obtain a Hilbert space $\check{\h}$ with the algebra $\check{\h}_0$ being a dense subspace of it. Consequently, we could define a representation of $\check{\h}_0$ acting on  $\check{\h}$ as follows
\begin{equation}
\check{\h}_0\ni \Psi\mapsto \hat{\Phi}\Psi:= {\Phi}\Psi\in \check{\h}_0
\label{repr-h-0}
\end{equation}
where $\Phi\in\check{\h}_0$. However, such a representation would not be  a $*$-rep\-re\-sen\-ta\-tion.   

To justify the last statement let us recall  that $\h$ is given by the scalar product satisfying \eqref{zero2} and consider a set ${\cal L}_0$ of independent loops consisting of at least two elements. Then there exist two {\em distinct} sets  ${\cal L}_a$ ($a=1,2$) of independent loops  such that ${\cal L}_0:={\cal L}_1\cup{\cal L}_2$. Given non-zero functions $\Psi_a\in \S(\Abar/\tl{\cal L}_a)$, the products $\Psi_1\Psi_2$ and $\Psi^2_1\Psi_2$ belong to $\S(\Abar/\tl{\cal L}_0)$. Assume now that the representation $\Psi\mapsto\hat{\Psi}$ given by \eqref{repr-h-0} is a $*$-representation and that the function $\Psi_1$ is real. Then the operator $\hat{\Psi}_1$ is symmetric on $\check{\h}_0$ (and on $\h_0$) and we have        
\begin{equation}
0\neq\scal{\Psi_1\Psi_2}{\Psi_1\Psi_2}=\scal{\hat{\Psi}_1\Psi_2}{\Psi_1\Psi_2}=\scal{\Psi_2}{\hat{\Psi}_1\Psi_1\Psi_2}=\scal{\Psi_2}{\Psi^2_1\Psi_2}=0,
\label{contr}
\end{equation}
where the last equation follows from \eqref{zero2}. 

The result means as well that the scalar product on $\h_0$ is not given by any positive functional defined on $\h_0$ (in particular by a measure on $\Abar$) i.e. if we assume that
\[
\scal{\Psi}{\Psi'}=\omega(\Psi^*\Psi'),
\]
where $\omega$ is a functional on $\h_0$ then the representation \eqref{repr-h-0}  is a $*$-rep\-re\-sen\-ta\-tion which leads to the above contradiction. Thus we conclude, in particular, that the spaces $\h$ and $\check{\h}$ (if the latter exists) are not of the form $L^2(\Abar,d\mu)$.   

Clearly, the source of the contradiction \eqref{contr} is $(i)$ the multiplication as the way we have defined the representation \eqref{repr-h-0} and $(ii)$ the orthogonality of the spaces $\S(\Abar/\tl{\cal L}_a)$ defined by {\em distinct} tame groups (Equation \eqref{zero2}). Therefore one can hope that it is possible to remove the problem by, either,
\begin{enumerate}
\item preserving the orthogonality and defining the representation of cylindrical functions in an other way,
\item giving up the orthogonality, while keeping the representation \eqref{repr-h-0},
\item giving up both the orthogonality and the multiplication as the way of defining the representation.
\end{enumerate}

Below we will present arguments which exclude the possibilties 1 and 2. By now we cannot exclude that the third possibility is a correct way to solve the problem, however, this possibility is not too promissing as working at the same time with a Hilbert space which does not satisfies \eqref{zero2} and with a representation diffrent from \eqref{repr-h-0} seems to be rather difficult. 

Thus the conclusion is that by now the space $\h$ and its modifications obtaining by giving up the orthogonality cannot be used in canonical quantization of any theory.
 
\subsection{Exclusion of the first possibility \label{ext}}

The argument which excludes the first possibility is given by Ashtekar \cite{a-notes}. Consider a $*$-algebra $\scripta$ containing two subalgebras $\scripta_1$ and $\scripta_2$  and Hilbert spaces $\h_i$, $i=0,1,2$. 

\begin{lm}
Suppose, $\pi$ is a $*$-representation\footnote{Let $L(\h)$ be a space of linear operators on a Hilbert space $\h$. We say that a map $\pi:\scripta\rightarrow L(\h)$ is a $*$-representation of $\scripta$ on the Hilbert space $\h$ if $(i)$ there exists a dense subspace ${\cal D}$ of $\h$ contained in $\bigcap_{\hat{a}\in\scripta}[\ D(\pi(\hat{a}))\cap D(\pi^*(\hat{a}))\ ]$, where $D(\pi(\hat{a}))$ denotes the domain of the operator $\pi(\hat{a})$ and $(ii)$ for every $\hat{a},\hat{b}\in\scripta$ and $\lambda\in\C$ the following conditions are satisfied on ${\cal D}$:
\begin{alignat*}{3}
\pi(\hat{a}+\hat{b})&=\pi(\hat{a})+\pi(\hat{b}),&\qquad\pi(\lambda\hat{a})&=\lambda\pi(\hat{a}),\\
\pi(\hat{a}\hat{b})&=\pi(\hat{a})\pi(\hat{b}),&\qquad \pi(\hat{a}^*)&=\pi^*(\hat{a}).
\end{alignat*} } 
of the algebra $\scripta$ on the orthogonal sum
\begin{equation}
\h_0\oplus\h_1\oplus\h_2
\label{hhil}
\end{equation}
such that, given $i,j\in\{1,2\}$,
\begin{enumerate}
\item the intersection of the domains of the operators $\{ \pi(a_i) \ | \ a_i\in \scripta_i\}$ contains a dense subspace ${\cal D}_j$ of $\h_j$,   
\item  for non-zero $i,j$ 
\[
\pi(a_i)v_j\in 
\begin{cases}
\h_i & \text{if $i=j$}\\
\h_0 & \text{if $i\neq j$} 
\end{cases}.
\]
where $a_i\in\scripta_i$ and $v_j\in {\cal D}_j$. Then for $i\neq j$ 
\[
\pi(\scripta_i){\cal D}_j=0.
\]
\end{enumerate} 
\label{lm-a}
\end{lm}

\noindent{\bf Proof.} Consider $\pi(a_1)v_2$, where $a_1\in\scripta_1$ and $v_2\in {\cal D}_2$. Then
\[
\scal{\pi(a_1)v_2}{\pi(a_1)v_2}=\scal{v_2}{\pi^*(a_1)\pi(a_1)v_2}=\scal{v_2}{\pi(a^*_1a_1)v_2}=0
\]  
as $v_2$ and $\pi(a^*_1a_1)v_2$  belong to the orthogonal subspaces of the Hilbert space \eqref{hhil}, what follows from the second assumption of the lemma. $\blacksquare$ 

To link the lemma with the discussed Hilbert space $\h$ it is enough to choose distinct tame hoop groups $\tilde{\cal L}_i$, $i=0,1,2$ such that  $\tilde{\cal L}_1$ and $\tilde{\cal L}_2$ are generated by distinct sets of independent loops, while $\tilde{\cal L}_0$ is generated by the union of the sets. Then 
\begin{gather*}
\scripta=\S(\Abar/\tl{\cal L}_0)\cup\S(\Abar/\tl{\cal L}_1)\cup \S(\Abar/\tl{\cal L}_2);\\
\scripta_i=\S(\Abar/\tl{\cal L}_i)={\cal D}_i, \ i=1,2;\\
\h_j=L^2(\Abar/\tl{\cal L}_j,d\mu_{\tl{\cal L}_j}), \ j=0,1,2.
\end{gather*}   


The second assumption of the lemma can seem to be unnatural at the first sight, but it is difficult to avoid it in the case of {\em non-compact} spaces $\{\Abar/\tl{\cal L}_j\}$ over which the Hilbert spaces under consideration are built --- the non-compactness forces us to assume that any 'reasonable' representation of cylindrical functions $\S(\Abar/\tl{\cal L}_i)$ acting on $L^2(\Abar/\tl{\cal L}_j,d\mu_{\tl{\cal L}_j})$ preserves the Hilbert space if $i=j$ and maps the space into the Hilbert space corresponding to the group $\tl{\cal L}_i\cup\tl{\cal L}_j$ if $i\neq j$. 

As concluded in \cite{a-notes}, the lemma means that any representation of cylindrical functions which satisfies the expectation written above leaves every Hilbert space $L^2(\Abar/\tl{\cal L},d\mu_{\tl{\cal L}})$ invariant. This property is also shared by the flux operators, thus the resulting representation of the elementary variables is {\em highly reducible} which is rather not acceptable from a physical perspective.       

\subsection{Exclusion of the second possibility}


Consider measure spaces $(\Omega_a,\mu_a)$ ($a=0,1,2$) such that $\Omega_0=\Omega_1\times\Omega_2$ (we do {\em not} assume that $\mu_0=\mu_1\times\mu_2$). Let a vector space $V$ be spanned by functions on $\Omega_0$ belonging to $\bigcup_{a=0}^2 L^2(\Omega_a,\mu_a)$.

\begin{lm}
Suppose that
\begin{enumerate}
\item there exists a scalar product $\scal{\cdot}{\cdot}$ on $V$ such that
\begin{enumerate}
\item if $f,f'\in L^2(\Omega_a,\mu_a)$, then:
\[
\scal{f}{f'}=\int_{\Omega_a}\bar{f} f'\,d\mu_a;
\]
\label{scal-int}
\item otherwise the product $\scal{\cdot}{\cdot}$ is only restricted by the following condition: for every real-valued function $F$ on  $\Omega_a$ and for functions $f,f'\in V$ such that $Ff,Ff'\in V$
\[
\scal{Ff}{f'}=\scal{f}{Ff'};
\]
\label{sym}
\end{enumerate}
\item  there exist measurable sets $U_1\subset\Omega_1$, $U_2\subset\Omega_2$ and $U_1\times U_2\subset\Omega_0$  such that
\[
\mu_1(U_1)<\mu_0(U_1\times U_2).
\]
\label{<}
\end{enumerate}
Then the scalar product  is not  positive definite.
\label{undef}
\end{lm}

\noindent{\bf Proof.} Let $f_1,f_2$ be characteristic functions of the subsets $U_1$ and $U_2$ respectively. Clearly, $F=f_2$, $f=f_1$ and $f'=f_1f_2$ satisfy the assumption \ref{sym}. Consider $f'':=f_1-f^2_2f_1\in V$. Then:
\begin{multline*}
\scal{f''}{f''}=\scal{f_1-f^2_2f_1}{f_1-f^2_2f_1}=\\=\scal{f_1}{f_1}+\scal{f^2_2 f_1}{f^2_2 f_1}-\scal{f_{1}}{f^2_2 f_1}-\scal{f^2_2 f_1}{f_{1}}=\\=\scal{f_1}{f_1}-\scal{f_2 f_1}{f_2 f_1}
=\int_{\Omega_1}f_1\,d\mu_{1}-\int_{\Omega_0}f_1f_2\,d\mu_0=\\=\mu_1(U_1)-\mu_0(U_1\times U_2)<0.
\end{multline*} 
$\blacksquare$ 

Let us set $\Omega_a=\Abar/\tl{\cal L}_a$ ($a=0,1,2$) and let $d\mu_a$ be an appropriate Haar measure $d\mu_{\tl{\cal L}_a}$ on $\Abar/\tl{\cal L}_a$. Notice now that the assumption \ref{scal-int} corresponds to Equation \eqref{integ} defining the scalar product of functions compatible with the same tame group, while the assumption \ref{sym} describes  what we mean by giving up the orthogonality. Finally, because
\[
d\mu_{\tl{\cal L}_0}=\frac{c_0}{c_1c_2}\,d\mu_{\tl{\cal L}_1}\times d\mu_{\tl{\cal L}_2},
\]
where $\{c_a\}$ are constants occurring in \eqref{haar-R}, and the spaces $\Abar/\tl{\cal L}_a$  are non-compact one can easily find sets $U_a$ satisfying the assumption \ref{<}.

We conclude that the Hilbert space $\h$ (and  $\check{\h}$) defined by the condition \eqref{zero2} and  the modifications of the space obtained by giving up the condition do not admit a $*$-representation of  functions on $\Abar$  defined by the multiplication map. 


\subsection{Other Hilbert spaces undergoing to the 'negative' results}

Although the space $\h$ is a very simple 'toy' model of a Hilbert space built over connections with a non-compact structure group the 'negative' results obtained above are valid in a more general case --- the contradiction \eqref{contr} and Lemmas \ref{lm-a} and \ref{undef} do not base at all on the assumption that the structure group is $\R$.  

In the literature there are presented two examples of Hilbert spaces  which suffer from the contradiction \eqref{contr} and are bound by Lemmas \ref{lm-a} and \ref{undef}; these are $(i)$ the Hilbert space $\h_{\rm FL}$ constructed by Freidel and Livine \cite{freidel} from gauge invariant cylindrical functions of connections with a {\em non-compact semi-simple} structure group and $(ii)$ the Hilbert space $\h_{\rm pr}$ of so called {\em projected  cylindrical functions} constructed by Livine \cite{liv}. 

\subsubsection{Hilbert space $\h_{\rm FL}$}

In fact, the Hilbert spaces $\h_{\rm FL}$ and $\h$ are built in a similar way and differences between the spaces are irrelevant to the reasoning which lead to the contradiction \eqref{contr} and the lemmas. The main differences are as follows: $(i)$ the cylindrical functions constituting $\h_{\rm FL}$  are compatible with graphs rather than with loops and $(ii)$ the scalar product on $\h_{\rm FL}$ of  two cylindrical functions compatible with the same graph $\Gamma$ is defined by means of a non-trivial measure $d\mu_\Gamma$ on a (non-compact) space\footnote{According to the notation used in the introduction $A_\Gamma\cong G^N/\sim$.} $A_\Gamma$ of 'gauge invariant parallel transports' along edges of the graph, while in the case of $\h$ we use a usual Haar measure on $\Abar/\tl{\cal L}\cong\R^n$. Consequently, 
\[
\h_{\rm FL}=\bigoplus^{\rm orth}_{\Gamma}L^2(A_\Gamma,d\mu_\Gamma),
\]
and the contradiction \eqref{contr} and Lemma \ref{lm-a} follow. On the other hand Lemma  \ref{undef} is general enough to be applied to appropriate modifications of $\h_{\rm FL}$.

\subsubsection{Hilbert space of projected cylindrical functions}

Neglecting some details of the construction described in \cite{liv}, one can say that the Hilbert space $\h_{\rm pr}(\g)$ of projected cylindrical functions compatible with a graph $\g$ is a subspace  of functions from the set $\A$ of $SO(1,3)$-connections over a three dimensional base manifold $\Sigma$ into the complex numbers. A function $\Psi:\A\rightarrow\C$ belongs to $\h_{\rm pr}(\g)$ if
\begin{enumerate}
\item $\Psi$  is of the form
\[
\Psi(A)=\psi(A(e_1),\ldots,A(e_N))
\] 
where $A(e_I)$ is the holonomy of the connection $A\in\A$ along the edge $e_I$ of the graph $\g$ and $\psi:SO(1,3)^N\rightarrow\C$;
\item $\psi$ satisfies the following gauge invariance condition: for every function $g:\Sigma\rightarrow SO(3)$ 
\[
\psi(A(e_1),\ldots,A(e_N))=\psi(\ g_{s,1} A(e_1) g_{t,1}^{-1},\ldots,g_{s,N}A(e_N)g^{-1}_{t,N}\ )
\]
where $g_{s,I}$ $(g_{t,I})$ is the value of the function $g$ in the source (the target) of the (oriented) edge $e_I$;
\item $\psi$ is a square integrable function with respect to the Haar measure $d\mu_{H}$ on $SO(1,3)^N$.  
\end{enumerate}      
Now, one defines a scalar products between two such functions $\Psi$ and $\Psi'$ as
\begin{equation}
\scal{\Psi}{\Psi'}=\int_{SO(1,3)^N}\bar{\psi}\psi'\,d\mu_{H}.
\label{scal-proj}
\end{equation}

Note that the requirement 2 does not contradict the requirement 3 because the group $SO(3)$ used to define the gauge invariance condition is {\em compact}. So $\psi$ is constatnt only along compact orbits in $SO(1,3)^N$, hence it can be square integrable with respect to the Haar measure and the scalar product \eqref{scal-proj} is well defined. The compactness of the gauge group is, however, the source of all problems known from the analysis of the Hilbert space $\h$ presented above: after reducing $SO(1,3)^N$ to the quotient space
\[
Q=SO(1,3)^N/\text{the gauge group}
\]   
we see that the scalar product \eqref{scal-proj} is defined by an intergral over the space $Q$ which is {\em non-compact}. Hence, while trying to 'glue' the spaces $\{\h_{\rm pr}(\g)\}$ labelled by all the graphs $\{\g\}$ in $\Sigma$ into one Hilbert space we immediately encounter obstacles as described by Lemmas \ref{lm-a} and \ref{undef}. 

\section{Summary}
 
Summarizing the results of the paper we conclude that an application of the canonical quantization in the form presented in \cite{cq-diff} to theories of connections with a non-compact structure group is still beyond our reach. The results obtained here suggest that slight modifications of the construction of the Hilbert space used in the compact case (like the ones presented here and in \cite{freidel}) may be insufficient to obtain a satisfactory result in the non-compact case. In particular, in the case of the Hilbert spaces introduced here and in \cite{freidel,liv} and, also, in the case of some modifications of the Hilbert spaces there exist obstacles which exclude a large class of representations of cylindrical functions  as $*$-representations. Moreover, at least in the case of the Hilbert space \eqref{ort-dec} it is not clear which cylindrical functions should be chosen as elementary variables.   


{\bf Acknowledgements:} I am grateful to Jerzy Lewandowski and Jerzy Kijowski for discussions and remarks and to Ahbay Ashtekar for his permission to quote some results from his unpublished Notes \cite{a-notes}. The Notes  also motivated me to formulate Lemma \ref{undef}. This work was supported in part by the Polish KBN grant  2 PO3B 068 23 and by NSF grant PHY 0090091.

\appendix

\section{Holonomy of an $\R$-connection \label{hol-app}}

Given a smooth connection $A$ on the bundle $P=\Sigma\times\R$  and an oriented path ${e}$ in $\Sigma$ a holonomy $H({e},A)$ denotes a parallel transport along ${e}$ defined by the connection $A$. To obtain an explicit formula describing the holonomy we will take advantage of the existence of the global trivialization of the bundle and express the connection $A$ by means of a differential one-form  on $\Sigma$ (denoted also by $A$) valued in the Lie algebra of the Lie group $\R$ (clearly, the Lie algebra is isomorphic to $\R$ as a vector space). Denoting by $\partial_x$ a left invariant vector field on $\R$ given by the Cartesian coordinate $(x)$ on $\R$ we obtain
\[
A=A_\mu(y) \, d y^\mu\ot \partial_x,
\]
where $(y^\mu)$ is a (local) coordinate frame on $\Sigma$. Let us parametrize the path,
\[
[a,b]\ni \tau \mapsto {e}_\tau \in \Sigma
\]
and denote by $\dot{{e}}_\tau$ the vector tangent to the curve $\tau\mapsto{e}_\tau$ at the point ${e}_\tau$. Consider now a differentiable map
\[
[a,b]\ni \tau \mapsto H(\tau)\in\R,
\] 
where $\R$ is understood as the Lie group, such that \cite{kobayashi}
\begin{equation}
[\dot{H}(\tau)\partial_x]_{H(\tau)}\cdot H^{-1}(\tau)=-\dot{{e}}_\tau \, \lrcorner \, A({e}_\tau) \ \ \text{and} \ \ H(a)=\mathbb{I}.
\label{hol-eq}
\end{equation}
In the above equation  $\cdot$ denotes the right action of the element $H^{-1}(\tau)\equiv -H(\tau)$ of the Lie group on the vector $[\dot{H}(\tau)\partial_x]_{H(\tau)}$ tangent to the group at the point $H(\tau)$; consequently the l.h.s. of the first of the above equations is a vector tangent to $\R$ in its neutral element $\mathbb{I}\equiv 0$. Let $x\mapsto f(x)$ be a differentiable function on $\R$. Then 
\[
[ \ [\dot{H}(\tau)\partial_x]_{H(\tau)}\cdot H^{-1}(\tau)\ ] \, f= \frac{d}{ds} f(H(\tau+s)-H(\tau))\big|_{s=0}=[\dot{H}(\tau)\partial_x]_\mathbb{I} \, f.
\] 
This means that Equation \eqref{hol-eq} can be written in the following form:
\[
\dot{H}(\tau)=-\dot{{e}}^\mu_\tau A_\mu({e}_\tau) \ \ \text{and} \ \ H(a)=0,
\]
hence
\[
H(\tau)=-\int_a^\tau \dot{{e}}^\mu_{\tau'}\, A_\mu({e}_{\tau'})\,d\tau'
\]
and
\begin{equation}
H({e},A)=-\int_a^b \dot{{e}}^\mu_{\tau'}\, A_\mu({e}_{\tau'})\,d\tau'=-\int_{e} A_\mu \, d y^\mu.
\label{hol-int}
\end{equation}

\section{The vector space $\h_0$ is not an algebra \label{h0-notal}}

To see this let us consider a set ${\cal L}_0=\{l_{0,1},l_{0,2},l_{0,3}\}$ of independent loops. Then the following sets defined for $a=1,2$
\begin{equation}
\begin{gathered}
{\cal L}_a:=\{l_a:=\prod_{\mu=1}^3 (l_{0,\mu})^{N_{a \mu}}\},\\
\ \ 
(N_{a\mu}):=
\begin{pmatrix}
1 & m_1 & n_1\\
n_2 & m_2  & 1
\end{pmatrix}, \ \ m_1,m_2,n_1,n_2\in \Z;
\end{gathered}
\label{lll}
\end{equation}
are (one-element) sets of independent loops. We assume also that the rank of the matrix $(N_{\mu a})$ is equal to 2. Now we are going to show by {\em reductio ad absurdum} that for some particular choice of the loops in ${\cal L}_0$ and the integers  defining the matrix $(N_{a\mu})$  the product of any two functions $\Psi_a\in\S(\Abar/\tl{\cal L}_a)$ does not belong to $\h_0$. 

Assume first that there exists a set  ${\cal L}'=\{l'_1,\ldots,l'_{n'}\}$ of independent loops such that 
\begin{equation}
\Psi_1\Psi_2\in \S(\Abar/\tl{\cal L}')\subset\h_0.
\label{pp-F}
\end{equation}
We can decompose the loops in  ${\cal L}'\cup {\cal L}_0$ in terms of loops ${\cal K}=\{k_1,\ldots,k_n\}$ as it is described in Subsection \ref{hg} whereby we obtain
\[
\tilde{l'}_{\nu}=\prod_{i=1}^{m}(\tilde{{k}}_i)^{M^{\prime}\!_{\nu i}}, \ \ \tilde{l}_{0,\mu}=\prod_{i=1}^{n}(\tilde{{k}}_i)^{M^0_{\mu i}},
\]  
where the matrices $(M^{\prime}\!_{\nu i})$ and $(M^0_{\mu i})$ are of the maximal rank.  Using the coordinates $(u_i=\bar{A}(\tl{k}_i))$ on $\Abar/\tl{\cal K}$ we can express functions $\Psi_a$ as follows
\[
\Psi_a(\bar{A})=\psi_a(\bar{A}(\tl{l}_a))=\psi_a(\sum_{i\mu} N_{a\mu}M^0_{\mu i } u_i),
\]      
where $\psi_a$ is a Schwarz function on $\R$, hence
\begin{multline}
(\Psi_1\Psi_2)(\bar{A})=\psi_1(\sum_{i\mu} N_{1\mu}M^0_{\mu i } u_i)\ \psi_2(\sum_{i\mu} N_{2\mu}M^0_{\mu i } u_i)\ =\\=\ \psi_{1,2}(\sum_{i\mu} N_{a\mu}M^0_{\mu i } u_i),
\label{psi-psi-0}
\end{multline}
where $\psi_{1,2}$ is a Schwarz function on $\R^2$. On the other hand, we have by virtue of the assumption \eqref{pp-F}
\[
(\Psi_1\Psi_2)(\bar{A})=\psi(\bar{A}(\tl{l}'_\nu))=\psi(\sum_i M'_{\nu i}u_i)
\]
for a function $\psi\in\S(\R^{n'})$. Thus
\begin{equation}
\psi(\sum_i M'_{\nu i}u_i) \ = \ \psi_{1,2}(\sum_{i\mu} N_{a\mu}M^0_{\mu i } u_i).
\label{psi-psi}
\end{equation}

It is clear that the rank of the matrix $(\sum_\mu N_{a\mu}M^0_{\mu i })$ is $2$. This means that the latter matrix defines a surjective linear map $N\circ M^0$ from $\R^n$ onto $\R^2$. Similarly, the matrix $(M'_{\nu i})$ defines a surjective linear map $M'$ from $\R^n$ onto $\R^{n'}$. Now Lemma \ref{3-stat} applied to Equation \eqref{psi-psi} implies immediately that $\ker M'=\ker(N\circ M^0)$. This fact has two consequences: $(i)$ the rank of $(M'_{\nu i})$ is  $2$ and therefore the set ${\cal L}'$ consists of {\em two} loops and $(ii)$ there exists an automorphism $Q$ of $\R^2$ such that $Q\circ M'=N\circ M^0$  or, equivalently,        
\[
\sum_{b} Q_{ab} M'_{b i} = \sum_{\mu} N_{a\mu} M^0_{\mu i}\in \Z.
\]       
Because suitably ordered columns of the matrix  $(M'_{bi})$ form a unit $(2\times 2)$-matrix the components of $(Q_{ab})$ belong to $\Z$. By means of  $\rho\equiv \rho_{\cal K}$ (i.e. the map defined by \eqref{ln}) we obtain 
\[
\sum_b Q_{ab}\, \rho(\tl{l}'_b)=\sum_\mu N_{a\mu}\, \rho(\tl{l}_{0,\mu})
\]
which is equivalent to
\begin{equation}
\prod_{b=1,2}(\tl{l}'_b)^{Q_{ab}}=\prod_{\mu=1,2,3}(\tl{l}_{0,\mu})^{N_{a \mu }}.
\label{l-lQN}
\end{equation}
The above equation describes a transformation between generators of the tame groups $\tl{\cal L}'$  and $\tl{\cal L}_0$ which are generated by, respectively, {\em two-element} and {\em three-element} sets of independent loops. We will show later on that there exist loops $\{{l}_{0,\mu}\}$ and the integers $m_1,m_2,n_1,n_2$ defining the matrix $(N_{a \mu})$ such that Equation \eqref{l-lQN} cannot hold. Because \eqref{l-lQN} is an implication of the assumption \eqref{pp-F} we conclude that for an appropriate choice of the loops  and the integers the product $\Psi_1\Psi_2$ does not belong to any space $\S(\Abar/\tl{\cal L})$.  

However, there might exist functions $\{\Psi'_b\in \S(\Abar/\tl{\cal L}'_b)\}$ such that
\begin{equation}
\Psi_1\Psi_2=\sum_b\Psi'_b.
\label{pp-sum}
\end{equation}
But then we can find a tame subgroup generated by $n$ loops which  contains all groups $\{\tl{\cal L}_0,\tl{\cal L}'_b\}$ under consideration. Then the functions $\{\Psi_1\Psi_2,\Psi'_b\}$ can be viewed as ones on $\R^n$ of the kind considered in Theorem \ref{lnzfunk} (see Equation \eqref{psi-psi-0}). Every such a function defines a linear subspace of $\R^n$ (see \eqref{Xf-0}). Assume now that a function $\Psi'_b$ and $\Psi_1\Psi_2$ define the same subspace of $\R^n$. Then using Lemma \ref{3-stat} 
we immediately conclude that $\Psi_1\Psi_2\in \S(\Abar/\tl{\cal L}'_b)$ which can be excluded as mentioned above. Then the function  $\Psi_1\Psi_2$ defines a subspace distinct from  ones given by $\{\Psi'_b\}$. By virtue of Theorem \ref{lnzfunk} we conclude that the functions on the both sides of \eqref{pp-sum} are linearly independent. Thus $\Psi_1\Psi_2$ does not belong to $\h_0$.    

Let us show finally that there exist loops $\{{l}_{0,\mu}\}$ and the integers $m_1,m_2,$ $n_1,n_2$ defining the matrix $(N_{a \mu})$ such that Equation \eqref{l-lQN} cannot be satisfied. 


Assume that $\{{l}_{0,\mu}\}$ are analytic loops such that the only common point of them is their base point $y$. Because the loops $\{l'_b\}$ are independent there  exist two analytic paths $\{{e}_b\}$  such that $(i)$ ${e}_b\subset l'_{b'}$ if and only if $b=b'$, $(ii)$ each of them as a part of the corresponding $l'_b$ is traced precisely once and $(iii)$ ${e}_1\cap{e}_2$ is an empty set. The analyticity of the loops $\{{l}_{0,\mu}\}$ means that each ${e}_b$ is contained precisely in one of the loops $\{{l}_{0,\mu}\}$. Moreover, $\{{l}_{0,\mu}\}$ is a set of {\em independent analytic} loops which means that every segment (including $e_b$) of each loop under consideration  is traced precisely once.

Suppose then that ${e}_1,{e}_2\subset{l}_{0,1}$. Then there exists a path ${e}_3\subset{l}_{0,3}$ which as an element of $l'_b$ can be traced $p_b\in \Z$ times (negative value of $p_b$ means that $e_3$ is traced in the direction opposite to the orientation of $l'_b$). Given $e_c$ ($c=1,2,3$), consider its 'characteristic' connection $A_c$, i.e. a connection in $\A$ such that $(i)$ $A_c(e_c)=1$ and $(ii)$ the support of the connection (one-form) does not contain any points of the loops $\{{l}_{0,\mu}\}$ and $\{l'_b\}$ except those of $e_c$.   
 
Now, acting on the both sides of \eqref{l-lQN} by the connections $\{A_c\}$ we obtain: 
\[
\sum_{b=1,2}Q_{ab}A_c(l'_b)=\sum_{\mu=1,2,3}N_{a\mu}A_c(l_{0,\mu}),
\]
that is
\[
Q_{a1} =N_{a1}=Q_{a2}, \ \ \text{and} \ \ Q_{a1}p_1+Q_{a2}p_2=N_{a3}.
\]
These equations can be satisfied only if $n_1n_2=1$. 

Of course, each path ${e}_b$ $(b=1,2)$ can belong to any loop ${l}_{0,\nu}$. Considering all possibilities we get a collection of conditions imposed on the integers $m_1,m_2,n_1,$ $n_2$ as, in fact, alternative necessary conditions for Equation \eqref{l-lQN} to be true:
\begin{gather*}
n_1n_2=1, \ \ n_2m_1=m_2, \ \ n_1m_2=m_1\\
\frac{n_1n_2-1}{n_2m_1-m_2},\frac{n_2n_1-1}{n_1m_2-m_1}\in \Z,\\
\frac{n_2m_1-m_2}{n_1n_2-1},\frac{n_1m_2-m_1}{n_2n_1-1}\in \Z.
\end{gather*}
Clearly, the following integers
\[
m_1=2, \ \ m_2=3, \ \ n_1=5, \ \ n_2=7
\]
$(i)$ do not satisfy any of the conditions,  $(ii)$ the matrix $(N_{a\mu})$ defined by them is of rank $2$ and  consequently \eqref{l-lQN} cannot hold for any two-element set ${\cal L}'=\{l'_{b}\}$ of independent loops. 

\section{Proof of Theorem \ref{thr-zero} \label{thr-zero-ap}}

We will show that for every $\tl{\cal L}_1\neq \tl{\cal L}_2$ there exist subsets $D_a\in \S(\Abar/\tl{\cal L}_a)$ ($a=1,2$) such that $D_a$ is a dense subspace of $L^2(\Abar/\tl{\cal L}_a,d\mu_{\tl{\cal L}_a})$ and for every pair of functions $\Psi_a\in D_a$ 
\begin{equation}
\scal{\Psi_1}{\Psi_2}=0,
\label{zero1}
\end{equation}
which is equivalent to \eqref{zero3}.

Let the tame group $\tl{\cal L}_a$ be generated by a set ${\cal L}_a=\{{l}_{a,1},\ldots,{l}_{a,n_a}\}$ of independent loops. Decompose the loops in ${\cal L}_1\cup{\cal L}_2$   in terms of independent loops ${\cal L}=\{{l}_1,\ldots,{l}_n\}$ as it is described in Subsection \ref{hg} (then $\tl{\cal L}_a$ is a subgroup of $\tl{\cal L}$). 

The function $\Psi_a\in\S(\Abar/\tl{\cal L}_a)$ is given by a function $\psi_a\in\S(\R^{n_a})$ according to \eqref{fun-cyl}. On the other hand $\Psi_a$ can be viewed as a cylindrical (but not Schwarz) function compatible with $\tl{\cal L}$, that is as an element of $\cyl(\Abar/\tl{\cal L})$. Therefore it is possible to express the function $\Psi_a$ by means of coordinates on $\Abar/\tl{\cal L}$. 

Following \eqref{A-xx} we define coordinates
\[
x_i=\bar{A}(\tl{l}_i), \ \ \ x_{a,\mu}=\bar{A}(\tl{l}_{a,\mu})
\]
on $\Abar/\tl{\cal L}$ and $\Abar/\tl{\cal L}_a$ respectively. By virtue of \eqref{AMA} we have 
\[
x_{a,\mu}=\sum_{i=1}^n M^a_{\mu i}x_i,
\]
where the matrix $(M^a_{\mu i})$ is given by the decomposition \eqref{matrix}. Therefore
\[
\Psi_a(\bar{A})=\psi_a(M^a_{\mu i}x_i).
\] 

On the other hand, given $a$, the matrix $(M^a_{\mu i})$ defines a subspace $V_a$ in $\R^n\cong\Abar/\tl{\cal L}$ (see Equations \eqref{rho-M} and \eqref{va}). Because $(\ker M^a)=V^\perp_a$ in the sense of the canonical scalar product on $\R^n\cong\Abar/\tl{\cal L}$ (see Equation \eqref{perp-V}) the space $V_a$ is isomorphic\footnote{The map $M^a$ given by the matrix $(M^a_{\mu i})$ is a surjective map from $\R^n\cong \Abar/\tl{\cal L}$ onto $\R^{n_a}\cong\Abar/\tl{\cal L}_a$ --- see \eqref{IMI}. The map restricted to $V_a\subset \R^n$ is defines the desired isomorphism between $V_a$ and $\Abar/\tl{\cal L}_a$.} to $\Abar/\tl{\cal L}_a$. Lemma \ref{l=va} guarantees that in our case $V_1\neq V_2$. Therefore we can treat the spaces $\Abar/\tl{\cal L}_1$ and $\Abar/\tl{\cal L}_2$ as {\em distinct} linear subspaces of $\Abar/\tl{\cal L}$. Thus without loss of generality we can assume that $\Abar/\tl{\cal L}_1\cong V_1\not\subset V_2\cong\Abar/\tl{\cal L}_2$.

Now it is easy to verify that there exists a coordinate frame $(u_i)$ on $\Abar/\tl{\cal L}$ such that  
\begin{enumerate}
\item $(u_i)$ are linear combinations of $(x_i)$,
\item coordinates $(u_1,\ldots,u_{n_1})$ as linear combinations of $(x_{1,\mu})$ are coordinates on $\Abar/\tl{\cal L}_1$,   
\item coordinates $(u_j,\ldots,u_{j+n_2})$ as linear combinations of $(x_{2,\mu})$ are coordinates on $\Abar/\tl{\cal L}_2$, 
\item $j>1$ i.e.  $u_1$ is not a coordinate on $\Abar/\tl{\cal L}_2$.   
\end{enumerate}    
Consequently, we have 
\[
\Psi_1(\bar{A})=\psi_1(u_1,\ldots,u_{n_1})\ \ \text{and} \ \ \Psi_2(\bar{A})=\psi_2(u_j,\ldots,u_{j+n_2}).
\]

The properties of the coordinates $(u_i)$ guarantee that there exists a linear combination (with real coefficients) $\hat{E}$ of flux operators \eqref{flux-op} operator such that
\[
\hat{E}\Psi_a=(\pi^*_{\tl{\cal L}}\circ {\cal I}^*_{\cal L}) (i\frac{\partial}{\partial u_1} \psi_a).
\] 
The assumptions of the theorem mean that $\hat{E}$ is self-adjoint with respect to the scalar product under consideration. Hence 
\begin{equation}
\scal{\hat{E}\Psi_1}{\Psi_2}=\scal{\Psi_1}{\hat{E}\Psi_2}=\scal{\Psi_1}{0}=0
\label{zero}
\end{equation}
We will show now that the above equation implies \eqref{zero1}, that is the thesis of the theorem.

Let us choose dense sets $D_a\in\S(\Abar/\tl{\cal L}_a)$ such that  $D_2=\S(\Abar/\tl{\cal L}_2)$ and  $D_1$ is a  space of finite linear combination of functions $\Psi_{m,\phi}$ $(m=0,1,2,\ldots)$ such that:    
\[
\Psi_{m,\phi}(\bar{A})=h_{m}(u_1)\phi(u_2,\ldots,u_{n_1})\in \S(\Abar/\tl{\cal L}_1),
\]
where $h_m:\R\rightarrow \R$ is a (normalized) Hermite function and\footnote{If $n_1=1$ let $\phi$ be just a constant.} $\phi\in\S(\R^{n_1-1})$. Obviously, $D_1$ is a dense subset in $L^2(\Abar/\tl{\cal L}_1,d\mu_{\tl{\cal L}_1})$.

The Hermite functions satisfy what follows:
\begin{equation}
\begin{gathered}
i\frac{\partial }{\partial u}h_m(u)=i\sqrt{\frac{m}{2}}h_{m-1}(u)-i\sqrt{\frac{m+1}{2}}h_{m+1}(u), \ m>0;\\ i\frac{\partial }{\partial u}h_0(u)=-i\frac{\sqrt{2}}{2}h_1(u)
\end{gathered}
\label{toz-poch}
\end{equation}

Let us fix an arbitrary non-zero functions $\phi\in \S(\R^{n_1-1})$ and $\Psi_2\in \S(\Abar/\tl{\cal L}_2)$ and assume that functions $\Psi_{m,\phi}$ and $\Psi_2$  are normalized with respect to the scalar product $\scal{\cdot}{\cdot}$ on $\h_0$. Setting $\Psi_{m,\phi}$ and $\Psi_2$ to (\ref{zero}) and applying the first equation of (\ref{toz-poch}) we get
\begin{equation}
0=\scal{\hat{E}\Psi_{m,\phi}}{\Psi_2}=i\sqrt{\frac{m}{2}}\scal{\Psi_{m-1,\phi}}{\Psi_2}-i\sqrt{\frac{m+1}{2}}\scal{\Psi_{m+1,\phi}}{\Psi_2}.
\label{herm-fun}
\end{equation}
Denoting
\[
I^+_{m}=\scal{\Psi_{2m,\phi}}{\Psi_2},\ \ \ I^-_{m}=\scal{\Psi_{2m+1,\phi}}{\Psi_2},\ \ \  m\geq 0,  
\]
we obtain from \eqref{herm-fun} recursive relations
\[
I^+_{m+1}=\sqrt{\frac{2m+1}{2m+2}}I^+_{m}, \ \ \ I^-_{m+1}=\sqrt{\frac{2m+2}{2m+3}}I^-_{m}.
\]
Now we will show that $I^+_{m}=0=I^-_{m}$ which is equivalent to vanishing of the scalar product between  functions in  $D_1$ and $D_2$ and therefore  will end the proof. 

Equation \eqref{zero} and the second equation of (\ref{toz-poch}) imply that
\[
0=\scal{\hat{E}\Psi_{0,\phi}}{\Psi_2}=-i\frac{\sqrt{2}}{2}\scal{\Psi_{1,\phi}}{\Psi_2}=-i\frac{\sqrt{2}}{2}I^-_{0},
\]
hence $I^-_{m}=0$. 

In order to show that $I^+_m=0$ 
let us consider a family of linear subspaces $\S_k\subset\S(\Abar/\tl{\cal L}_1)\oplus\S(\Abar/\tl{\cal L}_2)$ defined as
\begin{equation}
\S_k:=\text{span}\{\Psi_2,\Psi_{2,\phi},\Psi_{4,\phi},\ldots,\Psi_{2k,\phi}\}.
\label{span}
\end{equation}
The scalar product has to be positive definite  on $\h_0$, hence it does to be so on $\S_k$, in particular. A matrix $g_k$ of the product on $\S_k$ with respect to the basis (\ref{span}) is of the form:
\[
g_k=\left( 
\begin{array}{c|cccc}
1   & \bar{I}^+_1 &\bar{I}^+_2 & \ldots & \bar{I}^+_k\\ \hline
I^+_1 & 1         & 0          & \ldots & 0         \\
I^+_2 & 0	    & 1	 & \ldots & 0         \\  		 
\vdots & \vdots     & \vdots     & \ddots & 0         \\
I^+_k  & 0          &  0         &  0     & 1        
\end{array}
\right),
\]  
Its determinant has to be positive,
\[
\det g_k = (1-\sum_{l=1}^{k}|I^+_l|^2)=(1-|I^+_0|^2\sum_{l=1}^{k}w_l)>0,
\]
where $w_{l+1}=\frac{2l+1}{2l+2}w_l$ and $w_0=1$. The determinant is greater than zero if and only if
\[
|I^+_0|^2<\frac{1}{\sum_{l=1}^{k}w_l}.
\]
This requirement has to be satisfied for every $\S_{k}$, hence
\[
|I^+_0|^2\leq\lim_{k\rightarrow\infty}\frac{1}{\sum_{l=1}^{k}w_l}.
\]
Applying the Raabe's test it is easy to show that the series $(w_l)$ diverges to infinity:
\[
\lim_{l\rightarrow \infty}l(\frac{w_l}{w_{l+1}}-1)=\lim_{l\rightarrow \infty}l(\frac{2l+2}{2l+1}-1)=\frac{1}{2}<1.
\]  
It means that $I^+_0=0$ and therefore $I^+_m=0$. $\blacksquare$

\section{Other diffeomorphism invariant scalar products on $\h_0$ \label{non-orth}}

We are going now to define a diffeomorphism invariant, positive definite scalar product on $\h_0$ which does not satisfy the condition \eqref{zero2} i.e. such that the spaces $\S(\Abar/\tilde{\cal L}_1)$ and $\S(\Abar/\tilde{\cal L}_2)$ for some $\tilde{\cal L}_1\neq\tilde{\cal L}_2$ are not mutually orthogonal with respect to the product.   

Let us recall that $\mathbb{L}_y$ denotes the set of all piecewise analytic loops which originate and end at the point $y\in\Sigma$. Assume now that $(y^1,\ldots,y^d)$ $(d=\dim\Sigma)$ is a local analytic coordinate frame on $\Sigma$ such that $|y^i|\leq 2$ and values $(1,0,\ldots,0)$ of the coordinates correspond to the point $y$. Let a loop $l_0\in \mathbb{L}_y$ be given by
\[
\begin{cases}
y^1(t)=\cos t&\\
y^2(t)=\sin t&\\
y^i(t)=0&\text{for the remaining coordinates}
\end{cases}, \ \ t\in[0,2\pi].
\]         

Let $\tilde{\cal L}_0$ be a tame hoop group generated by the hoop $\tilde{l}_0$. As we have already stated, given an analytic diffeomorphism $\phi$ of $\Sigma$, the map \eqref{diff-hoop} can be understood as a map from $\cal HG$ onto $\cal HG$, i.e. $\widetilde{\phi(l_0)}$ can be naturally seen as a hoop based at the point $y$. Consequently, we denote by $\tl{\phi}(\tilde{\cal L}_0)$ a tame subgroup of $\cal HG$ generated by $\widetilde{\phi(l_0)}$. Let
\[
\tame_0\ :=\ \{\ \tl{\phi}(\tilde{\cal L}_0) \ | \ \phi\in{\rm Diff}^\omega(\Sigma)\ \},
\]  
where ${\rm Diff}^\omega(\Sigma)$ is the set of all analytic diffeomorphisms of $\Sigma$. Denote by $\tame$ the set of all tame subgroups of the hoop group $\cal HG$ and define 
\[
\tame_1:=\tame\setminus\tame_0.
\]     

The scalar product on $\h_0$ we are going to define will satisfy the condition
\begin{equation}
\scal{\cdot}{\cdot}_{\tilde{\cal L}_1\tilde{\cal L}_2}=0
\label{scal-a-0}
\end{equation}
if $\tilde{\cal L}_1\neq\tilde{\cal L}_2$ and either 
\begin{align*}
(i)& \ \ \tilde{\cal L}_1,\tilde{\cal L}_2\in\tame_1,\\
(ii)&  \ \ \tilde{\cal L}_1\in \tame_1 \ \text{and} \ \tilde{\cal L}_2\in\tame_0,\\
(iii)& \ \ \tilde{\cal L}_1\in\tame_0\ \text{and} \ \tilde{\cal L}_2\in\tame_1.
\end{align*}
However, if $\tilde{\cal L}_1,\tilde{\cal L}_2\in\tame_0$ then the spaces $\S(\Abar/\tilde{\cal L}_1)$ and  $\S(\Abar/\tilde{\cal L}_2)$ will not be mutually orthogonal with respect to the scalar product. 

To ensure this consider a linear space 
\[
\h'_0:=\bigoplus_{\tilde{\cal L}\in\tame_0}\S(\Abar/\tilde{\cal L})\subset\h_0. 
\]
By means of the Hermite functions $\{h_m\}_{m=0,1,\ldots}$ we define a basis of $\h'_0$:
\begin{equation}
\{ \ \Psi_{\tl{\cal L},m}:=(\pi^*_{\tl{\cal L}}\circ {\cal I}^*_{\cal L}) h_m \ | \ \tilde{\cal L}\in\tame_0, \ m=0,1,\ldots \  \}.
\label{basis}
\end{equation}
In the above formula, given $\tl{\cal L}$, $\cal L$ is a one-element set of independent loops generating the group  $\tl{\cal L}$. Clearly, $\cal L$ is determined by the loop $l_0$ and a diffeomorphism $\phi$. Thus one can choose either
\[
{\cal L}={\cal L^+}:=\{{e}^{-1}\circ \phi(l_0)\circ {e}\} \ \ \text{or} \ \ {\cal L}={\cal L^-}:=\{{e}^{-1}\circ(\phi(l_0))^{-1}\circ {e}\},
\]
where ${e}$ is an oriented piecewise analytic path originating at $y$ and ending at $\phi(y)$, and the loop $\phi(l_0)^{-1}$ is obtained from $\phi(l_0)$ by the change of its  orientation. We have consequently:
\begin{equation}
(\pi^*_{\tl{\cal L}}\circ {\cal I}^*_{\cal L^+}) h_m=(-1)^m\,(\pi^*_{\tl{\cal L}}\circ {\cal I}^*_{\cal L^-}) h_m.
\label{pm}
\end{equation}
Indeed, given $[\bar{A}]\in\Abar/\tl{\cal L}$,  
\[
{\cal I}_{\cal L^+}([\bar{A}])=\bar{A}({e}^{-1}\circ \phi(l_0)\circ {e})=-\bar{A}({e}^{-1}\circ \phi(l_0)^{-1}\circ {e})=-{\cal I}_{\cal L^-}([\bar{A}]).
\]
On the other hand functions $\{h_{2m}\}$ are even, and functions $\{h_{2m+1}\}$ are odd. Thus the basis \eqref{basis} is given modulo the factors $(-1)^m$. 

Now we can define a scalar product on $\h'_0$ as follows:
\begin{equation}
\scal{\Psi_{\tl{\cal L}',m'}}{\Psi_{\tl{\cal L},m}}:=
\begin{cases}
\delta_{m'm} & \text{if $\tl{\cal L}'=\tl{\cal L}$}\\
a & \text{if $\tl{\cal L}'\neq\tl{\cal L}$ and $m'=0=m$}\\
0 & \text{in the remaining cases}
\end{cases}
\label{scal-a}
\end{equation}
(notice that the resulting scalar product does not depend on the choice of the basis \eqref{basis}).

Every $\phi\in{\rm Diff}^\omega$ maps any basis \eqref{basis} onto a basis of the same form --- from \eqref{diff-cyl} and \eqref{pm} we have
\[
\bG_\phi [\Psi_{\tl{\cal L},m}]=\bG_{\phi} [\,(\pi^*_{\tl{\cal L}}\circ {\cal I}^*_{\cal L}) h_m\,]=(\pi^*_{\tl{\phi}(\tl{\cal L})}\circ {\cal I}^*_{\phi(\cal L)}) h_m=(\pm1)^m\Psi_{\tl{\phi}(\tl{\cal L}),m},
\] 
hence $\h'_0$ is preserved by all the diffeomorphisms. Moreover, the above expression shows that the scalar product \eqref{scal-a} on $\h'_0$ is diffeomorphism invariant. Consequently, the scalar product on $\h_0$ given by Equations \eqref{integ}, \eqref{scal-a-0} and \eqref{scal-a} is also diffeomorphism invariant. Now we have to show that the scalar product is positive definite.      

To do this let us notice that the space $\h_0$ equipped with the scalar product is of the form
\[
\h_0=\bigoplus^{\rm orth}_{\tl{\cal L}\in\tame_1}\S(\Abar/\tl{\cal L})\oplus^{\rm orth}{\cal K}_0\oplus^{\rm orth}{\cal K}_1,
\]
where
\begin{gather*}
{\cal K}_0:={\rm span}\{\ \Psi_{\tl{\cal L},0} \ | \ \tl{\cal L}\in\tame_0 \ \}\\
{\cal K}_1:={\rm span}\{\ \Psi_{\tl{\cal L},m} \ | \ \tl{\cal L}\in\tame_0 \ , m=1,2,\ldots \ \}
\end{gather*}
(clearly, $\h'_0={\cal K}_0\oplus^{\rm orth}{\cal K}_1$). Thus to prove that the scalar product under consideration is positive definite it is enough to show that it is so on ${\cal K}_0$. Because ${\cal K}_0$ is a space of {\em finite} linear combinations of  functions of the form $\Psi_{\tl{\cal L},0}$ we have to show only that for every $n=1,2\ldots$ the $(n\times n)$-matrix
\[
M_n=\left( 
\begin{array}{cccccc}
1 & a & a & a & \ldots & a \\ 
a & 1 & a & a & \ldots & a \\
a & a & 1 & a & \ldots & a \\  		 
a & a & a & 1 & \ldots & a \\  		 
\vdots & \vdots & \vdots  & \vdots & \ddots & \vdots   \\
a & a & a & a & \ldots & 1
\end{array}
\right)
\]     
is positive definite. This is true if determinants of all upper-left submatrices of $M_n$ are positive. Every such $(j\times j)$-submatrix $M_j$ of $M_n$ can be transformed into a matrix $M'_j$
\[
M'_j=\left( 
\begin{array}{ccccccc}
1 & a & a & a & \ldots &a& a \\ 
a-1& 1-a & 0 & 0& \ldots &0& 0 \\
0 & a-1 & 1 -a & 0 & \ldots &0& 0 \\  		 
0 & 0 & a -1& 1-a & \ldots &0& 0 \\  		 
\vdots & \vdots & \vdots  & \vdots & \ddots & \vdots& \vdots   \\
0 & 0 & 0 & 0& \ldots & a-1 & 1-a
\end{array}
\right)
\]
such that $\det M_j=\det M'_j$. By means of the method of induction we can easily show that
\[
\det (M'_j)=[(j-1)a+1](1-a)^{j-1}.
\]
Thus for $0<a<1$ the scalar product on $\h_0$ is positive definite.


\end{document}